\documentclass[12pt,a4paper]{article}

\usepackage{geometry}
\RequirePackage{amsmath}
\RequirePackage{mathrsfs}
\RequirePackage{amsfonts}
\RequirePackage{dsfont}
\RequirePackage{braket}
\RequirePackage{tabularx}
\usepackage{multirow}
\RequirePackage{nicefrac}
\RequirePackage{graphicx}
\RequirePackage{booktabs}
\RequirePackage{colortbl}
\RequirePackage{xcolor}
\RequirePackage[symbol]{footmisc}
\usepackage{mathtools}
\usepackage{comment}
\usepackage{blkarray}
\usepackage{stackengine}
\usepackage{float}
\usepackage{rotating}
\usepackage{hhline}

\def\AEF{A.E. Faraggi}

\def\IJMP#1#2#3{{\it Int.\ J.\ Mod.\ Phys.}\/ {\bf A#1} (#2) #3}

\def\EJP#1#2#3{{\it Eur.\ Phys.\ Jour.}\/ {\bf C#1} (#2) #3}

\def\JHEP#1#2#3{{\it JHEP}\/ {\bf #1} (#2) #3}
\def\NPB#1#2#3{{\it Nucl.\ Phys.}\/ {\bf B#1} (#2) #3}
\def\PLB#1#2#3{{\it Phys.\ Lett.}\/ {\bf B#1} (#2) #3}
\def\PRD#1#2#3{{\it Phys.\ Rev.}\/ {\bf D#1} (#2) #3}
\def\PRL#1#2#3{{\it Phys.\ Rev.\ Lett.}\/ {\bf #1} (#2) #3}

\def\etal{{\it et al\/}}

\def\beq{\begin{equation}}
\def\eeq{\end{equation}}
\def\beqn{\begin{eqnarray}}
\def\eeqn{\end{eqnarray}}
\def\ds{{$\tilde S$}}
\def\unahe{{${\overline{\rm NAHE}}$}}

\newcommand{\CC}[2]{C{#1\atopwithdelims[]#2}}

\newcommand{\ba}{\begin{eqnarray}}
\newcommand{\ea}{\end{eqnarray}}
\DeclareRobustCommand{\sqbinom}{\genfrac[]{0pt}{}}

\numberwithin{equation}{section}

\begin{document}
\begin{titlepage}
\samepage{
\setcounter{page}{1}
\rightline{}
\rightline{June 2020}

\vfill
\begin{center}
  {\Large \bf{Towards the Classification of Tachyon-Free Models \\ \medskip
      From Tachyonic Ten-Dimensional \\ \bigskip 
      Heterotic String Vacua}}

\vspace{1cm}
\vfill

{\large Alon E. Faraggi\footnote{E-mail address: alon.faraggi@liv.ac.uk}, 
         Viktor G. Matyas\footnote{E-mail address: viktor.matyas@liv.ac.uk}
 and Benjamin Percival\footnote{E-mail address: benjamin.percival@liv.ac.uk} 
}
\\

\vspace{1cm}

{\it Dept.\ of Mathematical Sciences, University of Liverpool, Liverpool
L69 7ZL, UK\\}

\vspace{.025in}
\end{center}

\vfill
\begin{abstract}
\noindent
Recently it was proposed that ten-dimensional tachyonic string vacua may serve
as starting points for the construction of viable four dimensional
phenomenological string models which are tachyon free. This is achieved
by projecting out the tachyons in the four-dimensional models
using projectors other than the projector which is utilised
in the supersymmetric models and those of the $SO(16)\times SO(16)$
heterotic string. We continue the exploration of this class of models
by developing systematic computerised tools for their classification,
the analysis of their tachyonic and
massless spectra, as well as analysis of their
partition functions and vacuum energy. We explore a randomly generated
space of $2\times10^9$ string vacua in this class and find that tachyon--free
models occur with $\sim 5\times 10^{-3}$ probability, and of those, 
phenomenologically inclined $SO(10)$ vacua with
$a_{00}=N_b^0-N_f^0=0$, {\it i.e.} equal number of fermionic and bosonic massless
states, occur with frequency $\sim 2\times 10^{-6}$.
Extracting
larger numbers of phenomenological vacua therefore requires
adaptation of fertility conditions that we discuss, and significantly increase
the frequency of tachyon--free models. Our results suggest that
spacetime supersymmetry may not be a necessary ingredient
in phenomenological string models, even at the Planck scale.

\end{abstract}

\smallskip}

\end{titlepage}

\section{Introduction}\label{intro}

String theory provides the tools to explore the connection of
quantum gravity with cosmological and particle physics 
data. For that purpose we need to construct toy models that mimic
the Standard Models of particle physics and cosmology.
The heterotic string \cite{hetecan}
is particularly appealing in this regard
since it gives rise to $SO(10)$ Grand Unified Theory structures
that are motivated by the Standard Model data. 
Since the mid--eighties the majority of string models studied
possessed $N=1$ spacetime supersymmetry. This is partly motivated by the
attractiveness of supersymmetry as an extension of the Standard Model.
It has a mathematically appealing symmetry structure that in its
local form requires spin 2 constituents. It alleviates the tension between
the electroweak and gravitational scales, and facilitates electroweak
symmetry breaking by dimensional transmutation and
this breaking is compatible with the observed data.
Furthermore, the observed Higgs mass is compatible with low
scale supersymmetry and indicates that the electroweak symmetry
breaking mechanism is perturbative. From the point of view of
string constructions, $N=1$ supersymmetry guarantees that the
vacuum is stable and simplifies the analysis of the spectra of
string models. It is clear, however, that supersymmetry has to be
broken at some scale, and despite its attractive features,
its realisation at low scales is not mandatory. 

Non--supersymmetric string models have also been of interest over the years. 
Those studies have mostly focused on analysing compactifications of the tachyon--free
ten dimensional $SO(16)\times SO(16)$ heterotic string
\cite{dh, itoyama, nonsusy, aafs, ADM, FR}. 
In addition to the supersymmetric, and non--supersymmetric
$SO(16)\times SO(16)$, models,
heterotic string theory gives rise to models that are
tachyonic in ten dimensions \cite{dh, kltclas, gv}.
Recently, it was argued that these
ten dimensional
tachyonic vacua can also serve as good starting points for constructing
tachyon free phenomenologically viable models \cite{spwsp, stable}. 
Whereas in the supersymmetric
and non--supersymmetric $SO(16)\times SO(16)$ heterotic string the
ten dimensional tachyons are projected out by the same projectors,
they can be projected in the four dimensional models by using
alternative projectors. Furthermore, a phenomenologically viable model
was presented in ref. \cite{stable}, and it was argued that in that
specific example all the moduli, aside from the dilaton, are fixed
perturbatively, and the dilaton may be fixed by the racetrack mechanism
\cite{racetrack}.
Hence, it was suggested that the model is stable, although it was acknowledged
that discussions of stability in non--supersymmetric string vacua are
at best speculative. Nevertheless, it illustrates the motivation
to analyse the ten dimensional tachyonic vacua, on par with their
supersymmetric and non--supersymmetric ten dimensional counterparts.
Extending the construction of phenomenological models to this
class of string compactifications may also shed light on some
of the outstanding issues in string phenomenology. 

Since the late eighties the heterotic string models in the free fermionic
formulation \cite{fff}
have provided a laboratory to study how the parameters of the Standard
Model are determined in a theory of quantum gravity
\cite{fsu5,fny, alr, slm, lrs, acfkr, su62,frs, slmclass, lrsclass, lrsfertile}.
These models correspond
to $Z_2\times Z_2$ toroidal orbifold compactifications with discrete Wilson
lines \cite{z2xz2}, and are related to compactifications on $Z_2$ orbifolds
of $K_3\times T_2$ manifolds. They give rise to a rich symmetry structure
from a purely mathematical point of view \cite{tw}.
Ultimately, this rich symmetry
structure may be reflected in the physical properties of these string
vacua, which is of future research interest. 
Three generation string models, with 
the $SO(10)$ GUT embedding of the Standard Model spectrum,
were constructed by following two routes. The first, the NAHE--based
models, use a common subset of boundary condition basis vectors,
the NAHE--set \cite{nahe}, which is extended by three or four additional
basis vectors. Using this method, models with different unbroken
$SO(10)$ subgroups were constructed \cite{fsu5, fny, alr, slm, lrs}.
The second route provides a powerful classification method of the
$Z_2\times Z_2$ toroidal orbifolds with different $SO(10)$ subgroups.
The initial classification method was developed for vacua with
unbroken $SO(10)$ subgroup \cite{fknr, fkr}
and subsequently extended to models
with Pati--Salam (PS) \cite{acfkr}; flipped $SU(5)$ (FSU5) \cite{frs};
Standard--like Model (SLM) \cite{slmclass};
and Left--Right Symmetric (LRS) \cite{lrsclass, lrsfertile}
$SO(10)$ subgroups. It led to the discovery of spinor--vector duality
in the space of $(2,0)$ string compactifications
\cite{fkr, cfkr, ffmt} and
the existence of exophobic string vacua \cite{acfkr}.
A NAHE--based tachyon--free
three generation Standard--like Model that descends
from a tachyonic ten-dimensional heterotic string vacuum was
constructed in ref. \cite{stable}.
The purpose of this paper is to initiate the systematic
classification of tachyon--free models that descends from a
particular class of tachyonic ten dimensional heterotic string
vacuum. Toward this end, we limit the classification to models
with unbroken $SO(10)$ subgroup, and extension to other
subgroups is left for future work. We focus on the systematic
analysis of the tachyonic and massless
sectors in the models; and the systematic
analysis of the partition function and the vacuum energy.
We remark that in contrast to the supersymmetric models,
systematic analysis of non--supersymmetric models
requires separate analysis of the spacetime fermionic and
bosonic sectors,
as they are no longer related by the spacetime supersymmetric map.
The computational time required is therefore doubled compared
to supersymmetric models, which motivates the development of novel
computational techniques \cite{fhpr}. In this regard, the direct analysis of
the partition function provides a powerful complementary tool.

Our paper is organised as follows: in Section \ref{freeferm} we
recap the main structure of the models that descend from
the ten dimensional tachyonic vacua.
The free fermionic classification method utilises a common
set of boundary condition basis vectors, which is presented
in Section \ref{freeferm}, and the enumeration
of the models is obtained by varying the one--loop Generalised
GSO projection coefficients. 
We discuss in sections \ref{freeferm} and
\ref{sectors} two generic maps that play important roles in our analysis,
the \ds--map (Section \ref{freeferm}), and
the ${\tilde x}$ map (Section \ref{sectors}).
In Section \ref{construction} we discuss the gauge symmetry arising
in our models and the sectors contributing to it. In sections
\ref{tsa} and \ref{sectors} we set up the tools for the systematic
analysis of the tachyonic and massless sectors of our models.
In Section \ref{PF} discuss the systematic analysis of the
partition function and the vacuum energy. In Section \ref{results}
we present the results of our classification. Section \ref{conclusion}
concludes our paper with a discussion of the results and outlook for
future research directions. 

\section{Ten Dimensional Vacua and the $\tilde{S}$-Map}\label{freeferm}
In the free fermionic construction, models are specified
in terms of boundary condition basis vectors and one--loop Generalised
GSO (GGSO) phases \cite{fff}. 
The $E_8\times E_8$ and $SO(16)\times SO(16)$
heterotic--models in ten dimensions are
defined in terms of a common set of basis vectors 
\ba
v_1={\mathds{1}}&=&\{\psi^\mu,\
\chi^{1,\dots,6} \ | \ \overline{\eta}^{1,2,3},
\overline{\psi}^{1,\dots,5},\overline{\phi}^{1,\dots,8}\},\nonumber\\
v_{2}=z_1&=&\{\overline{\psi}^{1,\dots,5},
              \overline{\eta}^{1,2,3} \},\nonumber\\
v_{3}=z_2&=&\{\overline{\phi}^{1,\dots,8}\},
\label{tendbasisvectors}
\ea 
where we adopted the common notation used in the
free fermionic models
\cite{fsu5, slm, alr, lrs, gkr, acfkr, su62, frs, 
slmclass, lrsclass, fknr, fkr, cfkr}. 
The basis vector $\mathds{1}$ is mandated by the modular invariance
consistency rules \cite{fff}, and produces a model with
$SO(32)$ gauge symmetry from the Neveu-Schwarz (NS) sector. 
The spacetime supersymmetry generator arises from the combination 
\beq
S={\mathds{1}}+z_1+z_2 = \{{\psi^\mu},\chi^{1,\dots,6}\}. 
\label{tendsvector}
\eeq
The choice of GGSO phase $\CC{z_1}{z_2}=\pm 1$
differentiates between 
the $SO(16)\times SO(16)$ or $E_8\times E_8$ heterotic strings 
in ten dimensions.
Eq. (\ref{tendsvector}) dictates that 
in ten dimensions the breaking of spacetime supersymmetry
is correlated with the breaking pattern 
$E_8\times E_8\rightarrow SO(16)\times SO(16)$. 
Equation (\ref{tendsvector}) does not hold in lower dimensions, 
and the two breakings are not correlated. On the other hand, these
vacua with broken and unbroken supersymmetry can be interpolated
\cite{interpol}. 

The tachyonic states in the $E_8\times E_8$ and $SO(16)\times SO(16)$ 
heterotic strings in ten dimensions are projected out. 
The would--be tachyons in these models are obtained from the Neveu-Schwarz (NS)
sector, by acting on the right--moving vacuum with a single fermionic 
oscillator
\beq
|0\rangle_L\otimes {\bar\phi}^a|0\rangle_R,
\label{untwistedtach}
\eeq
where in ten dimensions $a=1,\cdots, 16$.
The GSO projection induced by the $S$--vector projects 
out the untwisted tachyons,
producing tachyon free models in both cases. 
As discussed in refs. \cite{spwsp, stable}, obtaining the 
ten dimensional tachyonic vacua in the free fermionic formulation amounts
to the removal of the $S$--vector from the construction. 
The ten dimensional
configurations are obtained by substituting the $z_1$ basis vector
with $z_1=\{{\bar\phi}^{1,\cdots,~4}\}$ and adding similar $z_i$ 
basis vectors, with four periodic fermions,
and at most two overlapping.  
These vacua are connected by interpolations or orbifolds
along the lines of ref. \cite{gv}, and, in general, contain
tachyons in their spectrum.  

In the free fermionic formulation,
the four dimensional models that descend from the ten
dimensional tachyonic vacua amount to removing the vector $S$
from the set of basis vectors that are used to generate the models.
In four spacetime dimensions
the set $\{{\mathds1}, z_1,z_2\}$
produces a non--supersymmetric model with $SU(2)^6\times SO(12)\times 
E_8\times E_8$ or $SU(2)^6\times SO(12)\times SO(16)\times SO(16)$. 
An alternative to removing the $S$--vector from 
the construction is to augment it with periodic right--moving
fermions. A convenient choice is given by 
\beq
{\tilde S} = \{\psi^{1,2}, 
                \chi^{1,2},
                \chi^{3,4},
                \chi^{5,6} \ \vert \ {\bar\phi}^{3, \cdots,~6} \}\equiv~1~. 
\label{newS}
\eeq
In this case there are no massless gravitinos, and the untwisted 
tachyonic states 
\beq
|0\rangle_L\otimes {\bar\phi}^{3, \cdots,~6}|0\rangle_R 
\label{stildetachstates}
\eeq
are invariant under the ${\tilde S}$--vector projection.
These untwisted tachyons are those that descend 
from the ten dimensional vacuum, hence confirming that the 
model can be regarded as a compactification of a 
ten dimensional tachyonic vacuum.

We therefore observe a general map, which is induced by the exchange
\beq
S\longleftrightarrow {\tilde S},
\label{stildemap}
\eeq
in the construction of the heterotic string models that descend from
the ten dimensional tachyonic vacua. We refer to this map as the
${\tilde S}$--map. It was discussed and used in the construction
of the \unahe--based model in ref. \cite{stable}. We remark that the
\ds--map is reminiscent of the map used to induce the spinor--vector
duality in ref. \cite{fkr, ffmt}, in the sense that both utilise a block of
four periodic right--moving worldsheet fermions. We may term these
sorts of maps as modular maps, in the sense that they involve a block
of four periodic complex worldsheet fermions. We therefore have another instance
where such a modular map is reflected in the symmetry structure
of the string vacua. Be it the spacetime supersymmetry in the models
in which the $S$--basis vector is the supersymmetry spectral flow operator,
or in the spinor--vector dual models in which a similar spectral
flow operator operates in the observable $E_8$ sector and induces the
spinor--vector duality map \cite{fkr,ffmt}. Here, a similar operation is at play
in the four dimensional models inducing the transformation from
the supersymmetric (and non--supersymmetric) models that contain
the $S$--basis vector, to the non--supersymmetric models that contain
the \ds--basis vector. As discussed in ref. \cite{panos},
this may be a reflection of a larger symmetry structure that
underlies these models and string compactifications in general. 

\section{Non--Supersymmetric $SO(10)$ Models in 4D} \label{construction}
Let us now define the classification structure for the $SO(10)$ models we consider, which employ the \ds--map. The first ingredient we need is a set of basis
vectors that generate the space of $SO(10)$ \ds-models. We can choose the set
\begin{align}\label{basis}
\mathds{1}&=\{\psi^\mu,\
\chi^{1,\dots,6},y^{1,\dots,6}, w^{1,\dots,6}\ | \ \overline{y}^{1,\dots,6},\overline{w}^{1,\dots,6},
\overline{\psi}^{1,\dots,5},\overline{\eta}^{1,2,3},\overline{\phi}^{1,\dots,8}\},\nonumber\\
\tilde{S}&=\{{\psi^\mu},\chi^{1,\dots,6} \ | \ \overline{\phi}^{3,4,5,6}\},\nonumber\\
{e_i}&=\{y^{i},w^{i}\; | \; \overline{y}^{i},\overline{w}^{i}\}, 
\ \ \ i=1,...,6
\nonumber\\
{b_1}&=\{\psi^\mu,\chi^{12},y^{34},y^{56}\; | \; \overline{y}^{34},
\overline{y}^{56},\overline{\eta}^1,\overline{\psi}^{1,\dots,5}\},\\
{b_2}&=\{\psi^\mu,\chi^{34},y^{12},y^{56}\; | \; \overline{y}^{12},
\overline{y}^{56},\overline{\eta}^2,\overline{\psi}^{1,\dots,5}\},\nonumber\\
{b_3}&=\{\psi^\mu,\chi^{56},y^{12},y^{34}\; | \; \overline{y}^{12},
\overline{y}^{34},\overline{\eta}^3,\overline{\psi}^{1,\dots,5}\},\nonumber\\
z_1&=\{\overline{\phi}^{1,\dots,4}\},\nonumber
\nonumber
\end{align}
which is a similar basis set to \unahe=$\{\mathds{1},\tilde{S},b_1,b_2,b_3\}$ employed in \cite{stable}, except with the inclusion of $z_1$ to break the hidden gauge group and of $e_i$  to obtain all symmetric shifts  of the internal $\Gamma_{6,6}$ lattice. 
We note that the vector $b_3$ which spans the third twisted plane and 
facilitates the analysis of the obervable spinorial representations is typically formed as a linear combination in previous supersymmetric classifications \cite{fknr, fkr, acfkr, frs, slmclass, lrsclass, lrsfertile}. Furthermore we note the existence of a vector combination $z_2$ 
\beq \label{z2}
z_2=\mathds{1}+\sum_{i=1}^6 e_i +\sum_{k=1}^3 b_k+z_1=\{\bar{\phi}^{5,6,7,8}\}
\eeq 
in our models, which is typically its own basis vector
in previous classifications.

Models may then be defined through the specification of GGSO phases $\CC{v_i}{v_j}$, which for our $SO(10)$ models are 66 free phases with all others specified by modular invariance. Hence, the full space of models is of size $2^{66}\sim 10^{19.9}$ models. This is a notably enlarged space compared with the supersymmetric $SO(10)$ case where the requirement that the spectrum is supersymmetric
fixes some GGSO phases.

With a basis and a set of GGSO phases, we can construct the modular invariant 
Hilbert space $\mathcal{H}$ of states $\ket{S_\alpha}$ of the model 
through the one-loop GGSO projection such that
\begin{equation}
    \mathcal{H}=\bigoplus_{\alpha\in\Xi}\prod^{k}_{i=1}
\left\{ e^{i\pi v_i\cdot F_{\alpha}}\ket{S_\alpha}=\delta_{\alpha}
\CC{\alpha}{v_i}^*\ket{S_\alpha}\right\}
\end{equation}
where $\alpha$ is a sector formed as a linear combination of the basis vectors, $F_\alpha$ is the fermion number operator and $\delta_\alpha=1,-1$ 
is the spin-statistics index. 

The sectors in the model can be characterised according to the left and
right moving vacuum separately
\begin{align}
    M_L^2&=-\frac{1}{2}+\frac{\alpha_L \cdot\alpha_L}{8}+N_L\\
M_R^2 &=-1+\frac{\alpha_R \cdot\alpha_R}{8}+N_R \nonumber
\end{align}
where $N_L$ and $N_R$ are sums over left and right moving oscillators, 
respectively. Physical states must then additionally satisfy the
Virasoro matching condition: $M_L^2=M_R^2$, states not satisfying this
correspond to off-shell states. 

The untwisted sector gauge vector
bosons for this choice of basis vectors give rise to a gauge group 
\beq 
SO(10)\times U(1)_1\times U(1)_2\times U(1)_3\times SO(4)^4
\eeq 
where our desired GUT $SO(10)$ is generated by the spacetime vector bosons $\psi^\mu \bar{\psi}^a \bar{\psi}^b \ket{0}$, the $U(1)_{i=1,2,3}$ are those generated by the worldsheet currents $:\bar{\eta}^i\bar{\eta}^{i*}:$ and the $SO(4)^4$ is the hidden sector generated by spacetime vector bosons from the pairs of $\bar{\phi}^a$ with common boundary conditions for each basis vector: $\{\bar{\phi}^{1,2},\bar{\phi}^{3,4},\bar{\phi}^{5,6},\bar{\phi}^{7,8}\}$. 

The gauge group of a model may be enhanced by additional gauge bosons which may arise from the $z_1,z_2$ and $z_1+z_2$ sectors with appropriate oscillators, i.e.
\beq \label{enhancements}
\begin{Bmatrix}
\psi^\mu\ket{z_1}_L \otimes \{\bar{\lambda^i}\}\ket{z_1}_R \\ 
\psi^\mu\ket{z_2}_L \otimes \{\bar{\lambda^i}\}\ket{z_2}_R \\
\psi^\mu \ket{z_1+z_2}_L \otimes \ket{z_1+z_2}_R
\end{Bmatrix}
\eeq  
where $\bar{\lambda}^i$ are all possible right moving Neveu-Schwarz oscillators 

Whether these gauge bosons appear is model-dependent since it depends on
their survival under the GGSO projections. These enhancement sectors are
also present in the familiar supersymmetric classification set-ups used
in \cite{fknr,acfkr,frs,slmclass,lrsclass,lrsfertile}. However in those
cases there is also an observable enhancement from the vector
$x=\{\overline{\psi}^{1,...,5}, \overline{\eta}^{1,2,3}\}$,
which arises as a linear combination in these models.
If present, this vector induces the enhancement
$SO(10)\times U(1)\rightarrow E_6$, where the $U(1)=U(1)_1+U(1)_2+U(1)_3$
combination is typically anomalous \cite{cfua1},
unless such an enhancement is present.
This result was first discussed in the context of the NAHE models, where
including $x$ in the basis was shown to similarly produce $E_6$ GUT
models \cite{xmap}. We therefore can see that one effect of our
$\tilde{S}$ models with the basis (\ref{basis}) is to preclude the
possibility of an $E_6$ enhancement in these models.

From (\ref{enhancements}) we can deduce that enhancements of the observable $SO(10)$ gauge group may arise from the sectors
$\psi^\mu\{\bar{\psi}^a\}\ket{z_1},\psi^\mu\{\bar{\psi}^a\} \ket{z_2}$, $a=1,...,5$.
Interestingly, the sectors: $\ket{z_1},\ket{z_2}$ (with no oscillators) produce
level-matched tachyons with conformal weight $(-1/2,-1/2)$ and so the
appearance of these enhancements is correlated with the projection of
level-matched tachyons. The full analysis of the level-matched tachyonic
sectors is presented in the following section. 
\section{Tachyonic Sectors Analysis}\label{tsa}
Due to the absence of the supersymmetry generating vector $S$ in our
construction, analysing whether on-shell tachyons arise in the spectrum of
our models becomes paramount. On-shell tachyons will arise when 
\beq 
M_L^2=M_R^2<0,
\eeq 
which corresponds to left and right products of $\alpha_L \cdot\alpha_L=0,1,2,3$ and $\alpha_R \cdot\alpha_R=0,1,2,3,4,5,6,7$. The presence of such tachyonic sectors in the physical spectrum indicates the instability of the string vacuum. There are 126 of these sectors in our models which are summarised compactly in Table 1.
\begin{table}[!ht]
\centering
\begin{tabular}{|c|c|c|}
\hline
Mass Level&	Vectorials & Spinorials \\
\hline
$(-1/2,-1/2)$&$\{\bar{\lambda}^m\}\ket{0}$& \ $z_1,\ z_2$\\ 
\hline
$(-3/8,-3/8)$&$\{\bar{\lambda}^m\}e_i$&\ $e_i+z_1, \ e_i+z_2$\\
\hline
$(-1/4,-1/4)$& $\{\bar{\lambda}^m\}e_i+e_j$& $e_i+e_j+z_1, \ e_i+e_j+z_2$\\
\hline
$(-1/8,-1/8)$&$\{\bar{\lambda}^m\}e_i+e_j+e_k$ & $e_i+e_j+e_k+z_1, \  e_i+e_j+e_k+z_2$\\
\hline
\end{tabular}
\caption{\label{tachSectors} \emph{Level-matched tachyonic sectors and their
    mass level, where $i\neq j \neq k=1,...,6$} and $\bar{\lambda}^m$ is any right-moving complex fermion with NS boundary condition for the relevant tachyonic sector.}
\end{table}
We will find that models in which all 126 on-shell tachyons are projected by
the GGSO projections appear with probability $\sim 0.0054$ and so in our
classification we will throw away all but around 1 in 185 models.

In \cite{FR} a basis was chosen such that, rather than having the six
internal shift vectors $e_i$, the combinations $T_1=e_1+e_2$, $T_2=e_3+e_4$
and $T_3=e_5+e_6$ were employed. Such a grouping does not allow for sectors
to arise for all shifts in the internal space and, for example, means that
spinorial $\mathbf{16}/\overline{\mathbf{16}}$ sectors have a degeneracy
of 4 making 3 particle generations impossible once the $SO(10)$ group is
broken. However, choosing $T_{i=1,2,3}$ did have the advantage of restricting
the number of tachyonic sectors and allowing for a more simplified set-up to
perform an analysis of the structure of the 1-loop potential in
these models. 

Since finding models in which all on-shell tachyons are projected is of
utmost importance for all questions of stability of our string vacua we
will delineate the methodology used in our analysis. In order to perform
this analysis an efficient computer algorithm had to be developed which
could scan samples of $\mathcal{O}({10^9})$ or more for on-shell tachyons
within a reasonable computing time. The code we developed in python when
running in parallel across 64 cores could check a sample of $10^9$ models
for tachyons in approximately 12 hours. A more detailed analysis of how to
check whether our on-shell tachyons are projected is presented in the next section.
\subsection{Note on the proto-graviton}
Before we turn to the on-shell tachyon analysis we recall a general
result first discussed in \cite{D}, which states that every
non-supersymmetric string model necessarily contains off-shell tachyonic
states with conformal weight $(0,-1)$. In our models these are 
\beq \label{protograviton}
\psi^\mu\ket{0}_L \otimes \ket{0}_R,
\eeq 
which will appear in the spectrum independent of the GGSO coefficients.
We call such a state a ``proto-graviton'' and its guaranteed presence in
the string spectrum can be understood at the CFT level by noting that it appears
in the same Verma module as the graviton state which is always present in
the massless NS sector. 
\subsection{Tachyons of conformal weight $(-\frac{1}{2},-\frac{1}{2})$}
The first on-shell tachyons we will inspect are those with conformal weight
$(-\frac{1}{2},-\frac{1}{2})$. Firstly we have the aforementioned untwisted
tachyons (\ref{stildetachstates}) which are always projected since
$\binom{z_1}{NS}=\binom{z_2}{NS}=-\binom{b_i}{NS}=1$. Then there are then two spinorial tachyonic
sectors at this mass level: $z_{1}$ and $z_2$. The conditions for their
survival can be displayed as Tables \ref{tab1} and \ref{tab2}.

\newcommand\xrowht[2][0]{\addstackgap[.5\dimexpr#2\relax]{\vphantom{#1}}}
\begin{table}[!ht] \label{z1Tach}
\centering
\small
\setlength{\tabcolsep}{5pt}
\begin{tabular}{|c|c|c|c|c|c|c|c|c|c|c|}
\hline \xrowht[()]{10pt}
\textbf{Sector}&$\CC{z_1}{e_1}$&$\CC{z_1}{e_2}$&	$\CC{z_1}{e_3}$&$\CC{z_1}{e_4}$&$\CC{z_1}{e_5}$&	$\CC{z_1}{e_6}$&$\CC{z_1}{b_1}$&$\CC{z_1}{b_2}$&	$\CC{z_1}{b_3}$& $\CC{z_1}{z_2}$ \\ [0.75ex]
\hline
$z_1$&+&+&+&+&+&+&+&+&+&+ \\
\hline
\end{tabular}
\caption{\emph{Conditions on GGSO coefficients for survival of the on-shell tachyons $\ket{z_1}$}}
\label{tab1}
\end{table}
\begin{table}[!ht] \label{z2Tach}
\centering
\small
\setlength{\tabcolsep}{5pt}
\begin{tabular}{|c|c|c|c|c|c|c|c|c|c|c|}
\hline\xrowht[()]{10pt}
\textbf{Sector}&$\CC{z_2}{e_1}$&$\CC{z_2}{e_2}$&	$\CC{z_2}{e_3}$&$\CC{z_2}{e_4}$&$\CC{z_2}{e_5}$&	$\CC{z_2}{e_6}$&$\CC{z_2}{b_1}$&$\CC{z_2}{b_2}$&	$\CC{z_2}{b_3}$& $\CC{z_2}{z_1}$ \\
\hline
$z_2$&+&+&+&+&+&+&+&+&+&+ \\
\hline
\end{tabular}
\caption{\emph{Conditions on GGSO coefficients for survival of the
    on-shell tachyons $\ket{z_2}$}}
\label{tab2}
\end{table}
Which tells us that only when all 10 of the column phases are $+1$ do the
sectors remain in the spectrum. Interestingly, this has a bearing on the
existence of the gauge group enhancements mentioned in the previous section.
In particular, the only observable enhancements:
$\psi^\mu \ket{z_1}_L \otimes \bar{\psi}^a \ket{z_1}$ and
$\psi^\mu \ket{z_2}_L \otimes \bar{\psi}^a \ket{z_2}$ have the same
survival conditions as the $z_1,z_2$ tachyonic sectors.
Therefore we find that for our construction, there are no tachyon-free
models in which the $SO(10)$ is enhanced. This is evident in the
classification results shown in Table \ref{Statstable} of Section
\ref{results}. 
\subsection{Tachyons of conformal weight $(-\frac{3}{8},-\frac{3}{8})$}
Now moving up the mass levels to $(-\frac{3}{8},-\frac{3}{8})$, we have
vectorial tachyons from the 6 sectors:
$\{\bar{\lambda}^i\}\ket{e_i}$, $i=1,...,6$ and spinorial
tachyons from 12 sectors: $\ket{e_i+z_1}$ and $\ket{e_i+z_2}$.
To demonstrate how to check the survival of these sectors we
take the case of $\{\bar{\lambda}^i\}\ket{e_1}$, $\ket{e_1+z_1}$
and $\ket{e_1+z_2}$, which we show in the Tables \ref{tab11}, \ref{tab12} and \ref{tab13}.
The other cases with $e_{2,...,6}$ are much the same except for a
simple permutation of the projection phases.
\clearpage
\begin{table}[!ht] \label{33VectTachs}
\small
\setlength{\tabcolsep}{2pt}
\centering
\begin{tabular}{|c|c|c|c|c|c|c|c|c|c|c|}
\hline\xrowht[()]{10pt}
$\ket{e_1}$\textbf{ Oscillator}&$\CC{e_1}{\tilde{S}}$&$\CC{e_1}{e_2}$&$\CC{e_1}{e_3}$&	$\CC{e_1}{e_4}$&$\CC{e_1}{e_5}$&$\CC{e_1}{e_6}$&	$\CC{e_1}{b_1}$&$\CC{e_1}{\tilde{x}}$&$\CC{e_1}{z_1}$&	$\CC{e_1}{z_2}$ \\
\hline
$\{\bar{y}^2\}$&+&-&+&+&+&+&-&+&+&+ \\
\hline
$\{\bar{w}^2\}$&+&-&+&+&+&+&+&+&+&+ \\
\hline
$\{\bar{y}^3\}$&+&+&-&+&+&+&-&+&+&+ \\
\hline
$\{\bar{w}^3\}$&+&+&-&+&+&+&+&+&+&+ \\
\hline
$\{\bar{y}^4\}$&+&+&+&-&+&+&-&+&+&+ \\
\hline
$\{\bar{w}^4\}$&+&+&+&-&+&+&+&+&+&+ \\
\hline
$\{\bar{y}^5\}$&+&+&+&+&-&+&-&+&+&+ \\
\hline
$\{\bar{w}^5\}$&+&+&+&+&-&+&+&+&+&+ \\
\hline
$\{\bar{y}^6\}$&+&+&+&+&+&-&-&+&+&+ \\
\hline
$\{\bar{w}^6\}$&+&+&+&+&+&-&+&+&+&+ \\
\hline
$\{\bar{\psi}^{1/2/3/4/5 (*) }\}$&\multirow{2}{*}{+}&\multirow{2}{*}{+}&\multirow{2}{*}{+}&\multirow{2}{*}{+}&\multirow{2}{*}{+}&\multirow{2}{*}{+}&\multirow{2}{*}{-}&\multirow{2}{*}{-}&\multirow{2}{*}{+}&\multirow{2}{*}{+} \\
$/ \{\bar{\eta}^{1(*)} \}$ &&&&&&&&&&\\
\hline
$\{\bar{\eta}^{2,3(*)}\}$&+&+&+&+&+&+&+&-&+&+ \\
\hline
$\{\bar{\phi}^{1,2(*)}\}$&+&+&+&+&+&+&+&+&-&+ \\
\hline
$\{\bar{\phi}^{3,4(*)}\}$&-&+&+&+&+&+&+&+&-&+ \\
\hline
$\{\bar{\phi}^{5,6(*)}\}$&-&+&+&+&+&+&+&+&+&- \\
\hline
$\{\bar{\phi}^{7,8(*)}\}$&+&+&+&+&+&+&+&+&+&- \\
\hline
\end{tabular}
\caption{\emph{Conditions on GGSO coefficients for survival of the on-shell vectorial tachyons $\{\bar{\lambda}^i\}\ket{e_1}$. We have made use of the combination $\tilde{x}=b_1+b_2+b_3=\{\psi^\mu, \chi^{1,...,6} \ | \ \bar{\psi}^{1,2,3,4,5},\bar{\eta}^{1,2,3}\}$, which will be discussed more in the next section.}}
\label{tab11}
\end{table}
\begin{table}[!ht] \label{33z1Tachs}
\small 
\setlength{\tabcolsep}{2pt}
\centering
\begin{tabular}{|c|c|c|c|c|c|c|c|c|}
\hline\xrowht[()]{10pt}
\textbf{Sector}&$\CC{e_1+z_1}{e_2}$&$\CC{e_1+z_1}{e_3}$&	$\CC{e_1+z_1}{e_4}$&$\CC{e_1+z_1}{e_5}$&$\CC{e_1+z_1}{e_6}$&	$\CC{e_1+z_1}{b_1}$&$\CC{e_1+z_1}{\tilde{x}}$&	$\CC{e_1+z_1}{z_2}$ \\
\hline
$\ket{e_1+z_1}$&+&+&+&+&+&+&+&+ \\
\hline
\end{tabular}
\caption{\emph{Conditions on GGSO coefficients for survival of the on-shell tachyons $\ket{e_1+z_1}$}}
\label{tab12}
\end{table}
\begin{table}[!ht] \label{33z2Tachs}
\small 
\setlength{\tabcolsep}{2pt}
\centering
\begin{tabular}{|c|c|c|c|c|c|c|c|c|}
\hline\xrowht[()]{10pt}
\textbf{Sector}&$\CC{e_1+z_2}{e_2}$&$\CC{e_1+z_2}{e_3}$&	$\CC{e_1+z_2}{e_4}$&$\CC{e_1+z_2}{e_5}$&$\CC{e_1+z_2}{e_6}$&	$\CC{e_1+z_2}{b_1}$&$\CC{e_1+z_2}{\tilde{x}}$&	$\CC{e_1+z_2}{z_1}$ \\
\hline
$\ket{e_1+z_2}$&+&+&+&+&+&+&+&+ \\
\hline
\end{tabular}
\caption{\emph{Conditions on GGSO coefficients for survival of the on-shell tachyons $\ket{e_1+z_2}$}}
\label{tab13}
\end{table}
\subsection{Tachyons of conformal weight $(-\frac{1}{4},-\frac{1}{4})$}
Carrying on up the mass levels we have $(-\frac{1}{4},-\frac{1}{4})$ in which vectorial tachyons arise from 15 sectors: $\{\bar{\lambda}^i\}\ket{e_i+e_j}$, $i\neq j=1,...,6$ and spinorial tachyons arise from 30 sectors: $\ket{e_i+e_j+z_1}$ and $\ket{e_i+e_j+z_2}$. Again, we will present the conditions on the survival of $\{\bar{\lambda}^i\}\ket{e_1+e_2}$, $\ket{e_1+e_2+z_1}$ and $\ket{e_1+e_2+z_2}$ in Tables \ref{tab21}, \ref{tab22} and \ref{tab23} below and note that the other sectors with other $e_i$ combinations are easily obtainable from these.
\clearpage
\begin{table}[!ht] \label{22VectTachs}
\setlength{\tabcolsep}{1.25pt}
\renewcommand{\arraystretch}{1}
\footnotesize
\centering
\begin{tabular}{|c|c|c|c|c|c|c|c|c|c|}
\hline\xrowht[()]{10pt}
$\ket{e_1+e_2}$ &\multirow{2}{*}{$\CC{e_1+e_2}{\tilde{S}}$}&\multirow{2}{*}{$\CC{e_1+e_2}{e_3}$}&	\multirow{2}{*}{$\CC{e_1+e_2}{e_4}$}&\multirow{2}{*}{$\CC{e_1+e_2}{e_5}$}&\multirow{2}{*}{$\CC{e_1+e_2}{e_6}$}&\multirow{2}{*}{$\CC{e_1+e_2}{b_1}$}&\multirow{2}{*}{$\CC{e_1+e_2}{\tilde{x}}$}&\multirow{2}{*}{$\CC{e_1+e_2}{z_1}$}&	\multirow{2}{*}{$\CC{e_1+e_2}{z_2}$} \\
\textbf{Oscillators}&&&&&&&&&\\
\hline
$\{\bar{y}^3\}$&+&-&+&+&+&-&+&+&+ \\
\hline
$\{\bar{w}^3\}$&+&-&+&+&+&+&+&+&+ \\
\hline
$\{\bar{y}^4\}$&+&+&-&+&+&-&+&+&+ \\
\hline
$\{\bar{w}^4\}$&+&+&-&+&+&+&+&+&+ \\
\hline
$\{\bar{y}^5\}$&+&+&+&-&+&-&+&+&+ \\
\hline
$\{\bar{w}^5\}$&+&+&+&-&+&+&+&+&+ \\
\hline
$\{\bar{y}^6\}$&+&+&+&+&-&-&+&+&+ \\
\hline
$\{\bar{w}^6\}$&+&+&+&+&-&+&+&+&+ \\
\hline
$\{\bar{\psi}^{1/.../5 (*) }\}$&\multirow{2}{*}{+}&\multirow{2}{*}{+}&\multirow{2}{*}{+}&\multirow{2}{*}{+}&\multirow{2}{*}{+}&\multirow{2}{*}{-}&\multirow{2}{*}{-}&\multirow{2}{*}{+}&\multirow{2}{*}{+} \\
/$\{\bar{\eta}^{1(*)}\}$&&&&&&&&&\\
\hline
$\{\bar{\eta}^{2,3(*)}\}$&+&+&+&+&+&+&-&+&+ \\
\hline
$\{\bar{\phi}^{1,2(*)}\}$&+&+&+&+&+&+&+&-&+ \\
\hline
$\{\bar{\phi}^{3,4(*)}\}$&-&+&+&+&+&+&+&-&+ \\
\hline
$\{\bar{\phi}^{5,6(*)}\}$&-&+&+&+&+&+&+&+&- \\
\hline
$\{\bar{\phi}^{7,8(*)}\}$&+&+&+&+&+&+&+&+&- \\
\hline
\end{tabular}
\caption{\emph{Conditions on GGSO coefficients for survival of the on-shell vectorial tachyons $\{\bar{\lambda}^i\}\ket{e_1+e_2}$. We have made use of the combination $\tilde{x}=b_1+b_2+b_3=\{\psi^\mu, \chi^{1,...,6} \ | \ \bar{\psi}^{1,2,3,4,5},\bar{\eta}^{1,2,3}\}$, which will be discussed more in the next section.}} 
\label{tab21}
\end{table}
\begin{table}[!ht] \label{22z1Tachs}
\footnotesize
\centering
\setlength{\tabcolsep}{0.5pt}
\begin{tabular}{|c|c|c|c|c|c|c|c|}
\hline\xrowht[()]{10pt}
\textbf{Sector}&$\CC{e_1+e_2+z_1}{e_3}$&	$\CC{e_1+e_2+z_1}{e_4}$&$\CC{e_1+e_2+z_1}{e_5}$&$\CC{e_1+e_2+z_1}{e_6}$&	$\CC{e_1+e_2+z_1}{b_1}$&$\CC{e_1+e_2+z_1}{\tilde{x}}$&	$\CC{e_1+e_2+z_1}{z_2}$ \\
\hline
$\ket{e_1+e_2+z_1}$&+&+&+&+&+&+&+ \\
\hline
\end{tabular}
\caption{\emph{Conditions on GGSO coefficients for survival of the on-shell tachyons $\ket{e_1+e_2+z_1}$.}}
\label{tab22}
\end{table}
\begin{table}[!ht] \label{22z2Tachs}
\footnotesize
\setlength{\tabcolsep}{0.5pt}
\centering
\begin{tabular}{|c|c|c|c|c|c|c|c|}
\hline\xrowht[()]{10pt}
\textbf{Sector}&$\CC{e_1+e_2+z_2}{e_3}$&	$\CC{e_1+e_2+z_2}{e_4}$&$\CC{e_1+e_2+z_2}{e_5}$&$\CC{e_1+e_2+z_2}{e_6}$&	$\CC{e_1+e_2+z_2}{b_1}$&$\CC{e_1+e_2+z_2}{\tilde{x}}$&	$\CC{e_1+e_2+z_2}{z_1}$ \\
\hline
$\ket{e_1+e_2+z_2}$&+&+&+&+&+&+&+ \\
\hline
\end{tabular}
\caption{\emph{Conditions on GGSO coefficients for survival of the on-shell tachyons $\ket{e_1+e_2+z_2}$.}}
\label{tab23}
\end{table}
\normalsize

\subsection{Tachyons of conformal weight $(-\frac{1}{8},-\frac{1}{8})$}
The final mass level we obtain on-shell tachyons from is $(-\frac{1}{8},-\frac{1}{8})$, where vectorial tachyons arise from 20 sectors: $\{\bar{\lambda}^i\}\ket{e_i+e_j+e_k}$, $i\neq j\neq k=1,...,6$ and spinorial tachyons arise from 40 sectors: $\ket{e_i+e_j+e_k+z_1}$ and $\ket{e_i+e_j+e_k+z_2}$. We present the conditions on the survival of $\{\bar{\lambda}^i\}\ket{e_1+e_2+e_3}$, $\ket{e_1+e_2+e_3+z_1}$ and $\ket{e_1+e_2+e_3+z_2}$ in the Tables \ref{tab31}, \ref{tab32} and \ref{tab33} below and note again that the conditions for other sectors with other $e_i$ combinations are easily obtainable from these. \clearpage 

\begin{table}[!ht] \label{11VectTachs}
\centering
\footnotesize
\setlength{\tabcolsep}{0.5pt}
\begin{tabular}{|c|c|c|c|c|c|c|c|}
\hline\xrowht[()]{10pt}
$\ket{e_1+e_2+e_3}$&\multirow{2}{*}{$\CC{e_1+e_2+e_3}{\tilde{S}}$}&	\multirow{2}{*}{$\CC{e_1+e_2+e_3}{e_4}$}&\multirow{2}{*}{$\CC{e_1+e_2+e_3}{e_5}$}&\multirow{2}{*}{$\CC{e_1+e_2+e_3}{e_6}$}&\multirow{2}{*}{$\CC{e_1+e_2+e_3}{\tilde{x}}$}&\multirow{2}{*}{$\CC{e_1+e_2+e_3}{z_1}$}&	\multirow{2}{*}{$\CC{e_1+e_2+e_3}{z_2}$} \\
\textbf{ Oscillator }&&&&&&&\\
\hline
$\{\bar{y}^4/\bar{w}^4\}$&+&-&+&+&+&+&+ \\
\hline
$\{\bar{y}^5/\bar{w}^5\}$&+&+&-&+&+&+&+ \\
\hline
$\{\bar{y}^6/\bar{w}^6\}$&+&+&+&-&+&+&+ \\
\hline
$\{\bar{\psi}^{1/.../5 }\}$&\multirow{2}{*}{+}&\multirow{2}{*}{+}&\multirow{2}{*}{+}&\multirow{2}{*}{+}&\multirow{2}{*}{-}&\multirow{2}{*}{+}&\multirow{2}{*}{+} \\
$/\{\bar{\eta}^{1/2/3(*)}\}$&&&&&&&\\
\hline
$\{\bar{\phi}^{1,2(*)}\}$&+&+&+&+&+&-&+ \\
\hline
$\{\bar{\phi}^{3,4(*)}\}$&-&+&+&+&+&-&+ \\
\hline
$\{\bar{\phi}^{5,6(*)}\}$&-&+&+&+&+&+&- \\
\hline
$\{\bar{\phi}^{7,8(*)}\}$&+&+&+&+&+&+&- \\
\hline
\end{tabular}
\caption{\emph{Conditions on GGSO coefficients for survival of the on-shell vectorial tachyons $\{\bar{\lambda}^i\}\ket{e_1+e_2+e_3}$.}}
\label{tab31}
\end{table}
\begin{table}[!ht] \label{11z1Tachs}
\centering
\footnotesize 
\setlength{\tabcolsep}{1.5pt}
\begin{tabular}{|c|c|c|c|c|c|}
\hline\xrowht[()]{10pt}
\textbf{Sector}&$\CC{e_1+e_2+e_3+z_1}{e_4}$&$\CC{e_1+e_2+e_3+z_1}{e_5}$&$\CC{e_1+e_2+e_3+z_1}{e_6}$&$\CC{e_1+e_2+e_3+z_1}{\tilde{x}}$&	$\CC{e_1+e_2+e_3+z_1}{z_2}$ \\
\hline
$\ket{e_1+e_2+e_3+z_1}$&+&+&+&+&+ \\
\hline
\end{tabular}
\caption{\emph{Conditions on GGSO coefficients for survival of the on-shell  tachyons $\ket{e_1+e_2+e_3+z_1}$.}}
\label{tab32}
\end{table}
\begin{table}[!ht] \label{11z2Tachs}
\footnotesize 
\setlength{\tabcolsep}{1.5pt}
\centering
\begin{tabular}{|c|c|c|c|c|c|}
\hline\xrowht[()]{10pt}
\textbf{Sector}&$\CC{e_1+e_2+e_3+z_2}{e_4}$&$\CC{e_1+e_2+e_3+z_2}{e_5}$&$\CC{e_1+e_2+e_3+z_2}{e_6}$&$\CC{e_1+e_2+e_3+z_2}{\tilde{x}}$&	$\CC{e_1+e_2+e_3+z_2}{z_1}$ \\
\hline
$\ket{e_1+e_2+e_3+z_2}$&+&+&+&+&+ \\
\hline
\end{tabular}
\caption{\emph{Conditions on GGSO coefficients for survival of the on-shell  tachyons $\ket{e_1+e_2+e_3+z_2}$.}}
\label{tab33}
\end{table}

Using this structure of the conditions on the GGSO phases for the survival of tachyonic sectors at each mass level our computer algorithm runs through and checks whether any configuration of the phases that leaves the tachyon in the spectrum is satisfied. If none are satisfied then all 126 are projected and the model is retained for further analysis. 

Having dealt now with the $M_L^2=M_R^2<0$ level-matched sectors we turn our attention to the more familiar discussion of the structure of the massless sectors $M_L^2=M_R^2=0$ in the following section where we can discern the phenomenological features of our models.

\section{Massless Sectors} \label{sectors}
Now that we have a way to generate models free of on-shell tachyons,
we can turn our attention to the massless sectors and their representations.
Although some aspects of the massless spectrum look similar to the
supersymmetric case, the structure of our \ds -models are very different.
In particular, we can contrast our models with those in which supersymmetry
is spontaneously broken (by a GGSO phase) where in general some parts of
the spectrum remain supersymmetric. This was, for example, demonstrated
in \cite{FKP} in terms of invariant orbits of the partition function for
orbifold models with spontaneously broken supersymmetry. Similarly,
our models are very different than those of the broken supersymmertry models
discussed in \cite{aafs} where observable spinorial sectors of the models still exhibit 
a supersymmetric--like structure, {\it i.e.} in these sectors the bosonic and
fermionic states only differ by their charges under some $U(1)$ symmetries 
that are broken at a high scale. 
%

As we explore this new structure in the massless spectrum we will see
that the role of the $\tilde{S}$-map is of central importance. Further
to this, we will also uncover the importance of a vector combination
$\tilde{x}$ which induces another interesting map. Without the presence
of the supersymmetry generator $S$ we must also handle a number of extra
massless sectors which would not arise in supersymmetric setups due to
the GGSO projections induced by $S$.
\subsection{The Observable Sectors and the \ds \ and $\tilde{x}$-maps }
The chiral spinorial $\mathbf{16}/\overline{\mathbf{16}}$ representations
arise from the 48 sectors (16 from each orbifold plane)
\begin{eqnarray}\label{spins1}
B_{pqrs}^{(1)} &=&  b_1 + pe_3 + qe_4 + re_5 + se_6
\nonumber \\ 
&=& \{\psi^{\mu},\chi^{1,2},(1-p)y^3\bar{y}^3,
pw^3\bar{w}^3,(1-q)y^4\bar{y}^4,qw^4\bar{w}^4,
 \nonumber\\
& & ~~~ (1-r)y^5\bar{y}^5,rw^5\bar{w}^5,(1-s)y^6\bar{y}^6,
sw^6\bar{w}^6,\bar{\eta}^{1},\bar{\psi}^{1,\ldots,5}\}
 \\
B_{pqrs}^{(2)} &=&   b_2 + pe_1 + qe_2 + re_5 + se_6\nonumber \\
B_{pqrs}^{(3)} &=&   b_3 + pe_1 + qe_2 + re_3 + se_4\nonumber
\end{eqnarray}
where $p,q,r,s = 0,1$ account for all combinations of shift vectors of
the internal fermions $\{y^i,w^i \ | \ \bar{y}^i,\bar{w}^i\}$. As in
previous classifications, we can now write down generic algebraic
equations to determine the number $\mathbf{16}$ and $\overline{\mathbf{16}}$,
$N_{16}$ and $N_{\overline{16}}$, as a function of the GGSO coefficients.
To do this we first utilize the following projectors to determine which
of the 48 spinorial sectors survive
\begin{eqnarray}\nonumber
P^1_{pqrs} &=& \frac{1}{2^4}\prod_{i=1,2}\left(1-\CC{B^1_{pqrs}}{e_i}^* \right)\prod_{a=1,2}\left(1-\CC{B^1_{pqrs}}{z_a}^*\right)\\
P^2_{pqrs} &=& \frac{1}{2^4}\prod_{i=3,4}\left(1-\CC{
B^2_{pqrs}}{e_i}^*\right)\prod_{a=1,2}\left(1-\CC{B^2_{pqrs}}{z_a}^*\right)\\
\nonumber
P^3_{pqrs} &=& \frac{1}{2^4}\prod_{i=5,6}\left(1-\CC{B^3_{pqrs}}{e_i}^*\right)\prod_{a=1,2}\left(1-\CC{B^3_{pqrs}}{z_a}^*\right)
\end{eqnarray}
where, we recall that the vector $z_2=\{\overline{\phi}^{5,6,7,8}\}$ is
the combination defined in eq. (\ref{z2}).
Then we define the chirality phases
\begin{eqnarray}
X^1_{pqrs} &=& -\CC{B^1_{pqrs}}{b_2+(1-r)e_5+(1-s)e_6}^* \nonumber\\
X^2_{pqrs} &=& -\CC{B^2_{pqrs}}{b_1+(1-r)e_5+(1-s)e_6}^*\\
X^3_{pqrs} &=& -\CC{B^3_{pqrs}}{b_1+(1-r)e_3+(1-s)e_4}^*\nonumber
\end{eqnarray}
to determine whether a sector will give rise to a $\mathbf{16}$ or a
$\overline{\mathbf{16}}$. With these definitions we can write compact
expressions for $N_{16}$ and $N_{\overline{16}}$
\begin{align}\label{16s}
\begin{split}
N_{16} &= \frac{1}{2}\sum_{\substack{A=1,2,3 \\ p,q,r,s=0,1}} 
P_{pqrs}^A\left(1 + X^A_{pqrs}\right) \\
N_{\overline{16}} &= \frac{1}{2}\sum_{\substack{A=1,2,3 \\ p,q,r,s=0,1}} 
P_{pqrs}^A\left(1 - X^A_{pqrs}\right). 
\end{split}
\end{align}
Up to here these equations are familiar from previous supersymmetric
classifications. However there is a fundamental difference from the
supersymmetric case where $B^{1,2,3}$, along with all model sectors,
appear in supermultiplets with superpartners obtained through the
addition of $S$, which exchanges spacetime bosons with spacetime
fermions but leaves the gauge group representations unchanged.
In our set-up, the fermionic $B^{1,2,3}$ sectors have no such
bosonic sector counterparts. Indeed, the addition of our basis
vector $\tilde{S}$ would give rise to massive states with non-trivial 
representations under the hidden sector gauge group.
As mentioned above, we can also compare with the broken supersymmetry
models of \cite{aafs} where the bosonic counterparts of $B^{1,2,3}$
only differ from their fermionic superpartners by their charges under 
some $U(1)$ symmetries that are broken at a high scale.

A further new important feature of our construction is the inclusion
of the vector
\beq \label{xtilde}
\tilde{x}=b_1+b_2+b_3
\eeq 
which we name in analogy to the $x$--vector from the supersymmetric
classifications \cite{fkr, acfkr, frs, lrsclass, lrsfertile}.
We note that ${\tilde x}$ is the same as the vector $S+x$ which arises in
supersymmetric models.
In these models the states from the 
$x$--sector enhances the observable gauge symmetry from $SO(10)$ to $E_6$,
so $S+x$ arises when such an enhancement is present.
The vector $\tilde{x}$ is important in our models since it plays the role of
mapping between the observable spinorial and vectorial representations of $SO(10)$,
as well as a map between bosonic and fermionic states.
More specifically, the $\tilde x$--vector maps sectors that produce
spacetime fermions in the spinorial representation of $SO(10)$,
from which the Standard Model matter states are obtained, 
to sectors that produce spacetime bosons in its vectorial representation,
from which the Standard Model Higgs state is obtained. Thus, the
$\tilde x$--map induces simultaneously the fermion--boson map
of the $S$--vector, as well as the spinor--vector map of the $x$--vector. 
Without $S$ to provide the simple symmetry at each mass level between bosons
and fermions the question of the relationship between bosons and fermions
is unclear. It appears that the structure is controlled in some sense by
the \ds-map and the $\tilde{x}$-map taking us between mass levels as
both these maps often change the mass level of the sector they act on.

 
We also note that the $\tilde{x}$--sector also affects the observable
spectrum since its presence in the Hilbert space results in an
extra 4 $\mathbf{16}$'s and $\overline{\mathbf{16}}$'s of $SO(10)$.
The $\tilde{x}$--sector corresponds to the sector producing the
fermionic superpartners of the states from the $x$--sector, {\it i.e.}
$S+x$, which enhance the $SO(10)$ symmetry to $E_6$. The $\tilde{x}$--sector
therefore gives rise to the fermionic superpartners of the spacetime
vector bosons from the $x$--sector, which are in fact absent from the spectrum.
%
%

\subsection{Vectorial Sectors}
As mentioned above, the vector $\tilde{x}$ in (\ref{xtilde}) maps between the spinorial sectors $B^{1,2,3}_{pqrs}$ and vectorial sectors:
\begin{eqnarray}\label{vects}
V_{pqrs}^{(1)} &=&  B_{pqrs}^{(1)}+\tilde{x}\nonumber \\
&=&b_2+b_3 + pe_3 + qe_4 + re_5 + se_6
\nonumber \\ 
&=& \{\chi^{3,4,5,6},(1-p)y^3\bar{y}^3,
pw^3\bar{w}^3,(1-q)y^4\bar{y}^4,qw^4\bar{w}^4,
 \nonumber\\
& & ~~~ (1-r)y^5\bar{y}^5,rw^5\bar{w}^5,(1-s)y^6\bar{y}^6,
sw^6\bar{w}^6,\bar{\eta}^{2,3}\}
 \\
V_{pqrs}^{(2)} &=&   B_{pqrs}^{(2)}+\tilde{x}\nonumber \\
V_{pqrs}^{(3)} &=&   B_{pqrs}^{(3)}+\tilde{x}\nonumber
\end{eqnarray}
The observable vectorial $\mathbf{10}$ representations of $SO(10)$ arise when the right moving oscillator is a $\overline{\psi}^{a(*)}$, $a=1,...,5$. To determine the number of such observable vectorial sectors we use the projectors
\begin{eqnarray}\label{VectProjector}
\nonumber R^{(1)}_{pqrs} &=& \frac{1}{2^4}\prod_{i=1,2}\left(1 + \CC{e_i}{V^{(1)}_{pqrs}}
\right)
\prod_{a=1,2}\left[1+ \CC{z_a}{V^{(1)}_{pqrs}}\right)\\
R^{(2)}_{pqrs} &=& \frac{1}{2^4}\prod_{i=3,4}\left(1+\CC{e_i}{V^{(2)}_{pqrs}}
\right)
\prod_{a=1,2}\left(1 + \CC{z_a}{V^{(2)}_{pqrs}}\right)\\
\nonumber R^{(3)}_{pqrs} &=& \frac{1}{2^4}\prod_{i=5,6}\left(1+\CC{e_i}{V^{(3)}_{pqrs}}\right)
\prod_{a=1,2}\left(1 + \CC{z_a}{V^{(3)}_{pqrs}}\right).
\end{eqnarray}
Using these we can write the number of vectorial $\mathbf{10}$'s arising from these sectors as
\begin{eqnarray}
N_{10}=\sum_{\substack{A=1,2,3 \\ p,q,r,s=0,1 }}R^A_{pqrs}.
\end{eqnarray}
Further to these observable vectorials arising from $V^{1,2,3}$ there are the additional states arising for the other choices of oscillator $\bar{y}^i_{NS},\bar{w}^i_{NS},\bar{\phi}^{1,2},\bar{\phi}^{3,4},\bar{\phi}^{5,6},\bar{\phi}^{7,8}$, which only transform under the hidden group. 

In contrast to the supersymmetric case, our models come with additional vectorial sectors, which can give rise to states transforming under the observable gauge group as well as the hidden. 

Firstly we observe 4 additional sectors that can give rise to  vectorial states transforming under both the observable and the hidden or solely the hidden. These sectors are
\beq \label{addvects1}
\tilde{V}=\{\{\bar{\lambda}^i\}\ket{\tilde{S}},\{\bar{\lambda}^i\}\ket{\tilde{S}+z_1},\{\bar{\lambda}^i\}\ket{\tilde{S}+z_2}, \{\bar{\lambda}^i\}\ket{\tilde{S}+z_1+z_2}\} 
\eeq 
which are spacetime fermions. 
There are two cases to distinguish when one of these sectors is present:
\begin{itemize}
\item $\{\bar{y}^i/\bar{w}^i\} \ket{\tilde{V}}$
  which are charged under the hidden sector only.
\item $\{ \bar{\psi}^{1,...,5},\bar{\eta}^{1,2,3},\bar{\phi}_{NS}\}
  \ket{\tilde{V}}$ with $\bar{\phi}_{NS}$ being the four
  Neveu-Schwarz oscillators such that $\bar{\phi}_{NS} \cap \tilde{V} =
  \emptyset$. 
  These transform in mixed
  representations of the observable and hidden sectors which means
  we should analyse them further. We realise that the
  condition for one of these to remain in the spectrum is:
    \beq 
    \CC{\tilde{V}}{e_i}=-1, \ \ \ \ \forall \ i\in \{1,2,3,4,5,6\}
    \eeq 
    for one of the $\tilde{V}$. In ref. \cite{stable} it was suggested that
    such states appearing in these models may be instrumental in implementing
    electroweak symmetry breaking by hidden sector condensates.
%
\end{itemize}
Similar to the $\tilde x$--sector, it is interesting to compare the \ds--sector with the
$S$--sector in supersymmetric models. 
%
The $S$--sector in the supersymmetric models produces the spacetime
fermionic superpartners of the states from the
NS--sector, {\it i.e.} it gives rise to the gauginos. The \ds--sector
gives rise to spacetime fermions that could transform as, {\it e.g.} electroweak
doublets and triplets, but also transforms as doublets of the hidden
gauge group, due to the \ds--map noted in Section \ref{freeferm}.
In this respect the \ds--models exhibit a sort of split supersymmetry,
in the sense that the states from the sectors $B_{1,2,3}$ are massive,
but the sector that produces the would--be gauginos, {\it i.e.} \ds,
still produces massless states transforming under the observable gauge
symmetry. It will be of interest to explore
how this phenomenon affects the phenomenological
characteristics of the models. 

Finally, there are further vectorials that may be observable or
hidden arising from the 15 sectors
\beq \label{addvects2}
\gamma^{k=1,...,15}=\{\bar{\lambda}^i \} \ket{e_i+e_j+e_k+e_l}
\eeq 
for $i\neq j \neq k \neq l =1,...,6$. 

We note that these sectors can give rise to vectorial $\mathbf{10}$'s
when the oscillators $\bar{\psi}^a$, $a=1,...,5$, are present.
In this case the projector is
\beq 
P_{\gamma^k}=\frac{1}{2^5}\prod_{i=m,n}\left(1+\CC{\gamma^k}{e_i} \right)
\prod_{a=1,2}\left(1 + \CC{\gamma^k}{z_a} 
\right)\left(1-\CC{\gamma^k}{\tilde{x}}\right)
\eeq 
where $m\neq n \neq i \neq j \neq k \neq l$. We can count the number of
such sectors through the expression
\beq 
N_\gamma^{\{\bar{\psi},\bar{\eta}\}}=\sum_{k=1}^{15}P_{\gamma^k}.
\eeq 
These additional vectorials can evidently play a role in the
phenomenology of our models, so their couplings and charge contributions
must be considered carefully for specific models. We can note that
$\gamma^k$ will not couple at leading order to the observable spinorial 
representations due to their additional charges, and
so at leading order the only vectorial 10 representations to generate realistic 
Standard Model fermion mass spectrum, remain those from
$V^{1,2,3}$.  
\subsection{Hidden Sectors}
We find that there are a relatively large number of hidden massless
sectors in our model, which is another effect of the  \ds -map we
have chosen, since its right moving complex fermions generate
representations of the hidden group. 

Firstly, we can identify 96 spinorial sectors that give rise to
spacetime bosons arising through the addition of $z_1$ or
$z_2$ onto the vectorial sectors $V^{1,2,3}$
\begin{eqnarray}\label{spins2}
H_{pqrs}^{(1)} &=&  V^{(1)}_{pqrs}+z_1
\nonumber \\ 
H_{pqrs}^{(2)} &=&   V^{(2)}_{pqrs}+z_1 \nonumber\\
H_{pqrs}^{(3)} &=&   V^{(3)}_{pqrs}+z_1\\
H_{pqrs}^{(4)} &=&  V^{(1)}_{pqrs}+z_2
\nonumber \\ 
H_{pqrs}^{(5)} &=&   V^{(2)}_{pqrs}+z_2 \nonumber\\
H_{pqrs}^{(6)} &=&   V^{(3)}_{pqrs}+z_2\nonumber
\end{eqnarray}
which evidently transform under the hidden $SO(4)^4$ only. 

A further four groups of 48 sectors are generated through the
addition of the combinations $\{\tilde{S},\tilde{S}+z_1,
\tilde{S}+z_2,\tilde{S}+z_1+z_2\}$ which give rise to spacetime
fermionic hidden sectors: 
\begin{eqnarray}\label{spins3}
H_{pqrs}^{(7)} &=&  \tilde{S}+V^{(1)}_{pqrs}
\nonumber \\ 
H_{pqrs}^{(8)} &=&   \tilde{S}+V^{(2)}_{pqrs} \nonumber\\
H_{pqrs}^{(9)} &=&  \tilde{S}+ V^{(3)}_{pqrs}\nonumber\\
H_{pqrs}^{(10)} &=&  \tilde{S}+V^{(1)}_{pqrs}+z_1
\nonumber \\ 
H_{pqrs}^{(11)} &=&   \tilde{S}+V^{(2)}_{pqrs}+z_1 \nonumber\\
H_{pqrs}^{(12)} &=&   \tilde{S}+V^{(3)}_{pqrs}+z_1\nonumber\\
H_{pqrs}^{(13)} &=&  \tilde{S}+V^{(1)}_{pqrs}+z_2 \\ 
H_{pqrs}^{(14)} &=&   \tilde{S}+V^{(2)}_{pqrs}+z_2 \nonumber\\
H_{pqrs}^{(15)} &=&   \tilde{S}+V^{(3)}_{pqrs}+z_2\nonumber\\
H_{pqrs}^{(16)} &=&  \tilde{S}+V^{(1)}_{pqrs}+z_1+z_2
\nonumber \\ 
H_{pqrs}^{(17)} &=&   \tilde{S}+V^{(2)}_{pqrs}+z_1+z_2 \nonumber\\
H_{pqrs}^{(18)} &=&   \tilde{S}+V^{(3)}_{pqrs}+z_1+z_2\nonumber
\end{eqnarray}
Essentially we see that by adding on the combinations:
$\tilde{h}_n=\{z_1,z_2,\tilde{S},\tilde{S}+z_1,
\tilde{S}+z_2,\tilde{S}+z_1+z_2\}$ we generate the
6 ways of having 2 doublet representations of the four hidden $SO(4)$ groups. 
Knowing the number of hidden sectors will mainly be useful when
looking at the size of massless coefficient in the $q$-expansion of
the partition function, which is equivalent to a counting of the number
of massless states. We will return to this
in Section \ref{PF}.

There are additional hidden sectors, on top of those counted by $N_H$,
that don't live on the orbifold planes. These 30 sectors are:
\begin{eqnarray}\label{spins4}
\delta^{1,...,30}=\begin{cases}
e_i+e_j+e_k+e_l+z_1\\ 
e_i+e_j+e_k+e_l+z_2
\end{cases}
\end{eqnarray}
for $i\neq j \neq k \neq l =1,...,6$. Similar to
(\ref{addvects1}), (\ref{addvects2}) these are examples of sectors
which are  not found in supersymmetric models since the $S$--vector
would project them out. Again, in order to evaluate the massless
contribution to the $q$-expansion we will need to count the states
arising from these sectors.

\section{Partition Function and Cosmological Constant} \label{PF}

The partition function of string models encapsulates all the 
information one knows about its structure, symmetries and spectrum.
Thus to fully understand our model it is essential to get a handle on
the calculation and form of its partition function.
The analysis of the partition function is particularly instrumental
in non--supersymmetric constructions, since it gives a complementary
tool to count the total number of massless states, and its integration
over the fundamental domain correspond to the cosmological constant.

For free fermionic models, the partition function
receives contributions from both the
fermionic and bosonic coordinates. That is the partition function can be
split into $Z=Z_BZ_F$. Since the fermions are free, their contribution to
the partition function is purely determined in terms of their boundary
conditions on the world-sheet torus, hence we can write
\begin{equation}
  Z = Z_B \,\sum_{Sp.Str.} \CC{\alpha}{\beta} \prod_{f} Z
  \sqbinom{\alpha(f)}{\beta(f)},
  \label{Z}
\end{equation}
where $\alpha$ and $\beta$ represent the boundary conditions,
the sum is over all choices of spin structure and the product is
over all fermions. The GGSO coefficients $\CC{\alpha}{\beta}$ are chosen
so that $Z_F$ is modular invariant. The $Z[\alpha(f),\beta(f)]$
terms are given as
\begin{align}
\begin{split}
    Z \sqbinom{1}{1} =&\, \sqrt{\frac{\vartheta_1}{\eta}},\qquad
    Z \sqbinom{1}{0} =\, \sqrt{\frac{\vartheta_2}{\eta}},\\
    Z \sqbinom{0}{0} =&\, \sqrt{\frac{\vartheta_3}{\eta},}\qquad
    Z \sqbinom{0}{1} =\, \sqrt{\frac{\vartheta_4}{\eta}},
\end{split}
\end{align}
where  $\vartheta_i$ and $\eta$ are the Jacobi theta functions and the
Dedekind eta function respectively. The bosonic term $Z_B$ comes from
$\partial X$, the bosonic superpartners of the $\psi$. Their
contribution to the partition function in four dimensions is given by 
\begin{equation}
  Z_B = \frac{1}{\tau_2} \frac{1}{\eta^2\bar{\eta}^2},
  \label{Z_B}
\end{equation}
where $\tau_2$ is the the imaginary part of the modular parameter.

The partition function (\ref{Z})  is a function of the modular parameter
$\tau = \tau_1 + i\tau_2$, which parametrises the one-loop world-sheet torus.
Thus, to get a numerical value from the one-loop partition function,
one has to sum over all the inequivalent tori, i.e. all values of
$\tau$ that give tori that are not related by modular transformations.
This region of the complex plane is referred to as the fundamental domain
of the modular group and is denoted $\mathcal{F}\subset\mathbb{C}$, with
$$ \mathcal{F} = \{\tau\in\mathbb{C}\,|\,|\tau|^2>1 \;\land\;|\tau_1|<1/2\}. $$
The full partition function therefore is given by the integral of (\ref{Z})
over this domain, specifically
\begin{equation}
    Z = \int_\mathcal{F}\frac{d^2\tau}{\tau_2^2}\, Z_B  \sum_{Sp.Str.} \CC{\alpha}{\beta} \prod_{f} Z \sqbinom{\alpha(f)}{\beta(f)},
    \label{ZInt}
\end{equation}
where $d^2\tau/\tau_2^2$ is the modular invariant measure. The expression (\ref{ZInt}) specifically represents the one-loop vacuum energy of our theory and so we may refer to it as the cosmological constant $\Lambda$. 

\begin{figure}[t]
\centering
\includegraphics[width=0.8\linewidth]{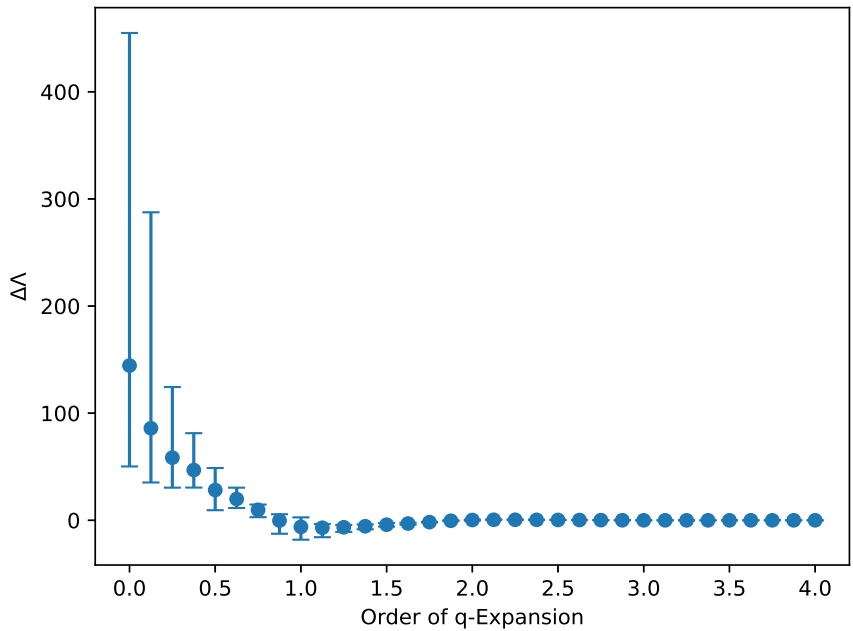}
\caption{\emph{The convergence of $\Lambda$ order-by-order in the q-expansion, where $\Delta \Lambda$ is the difference between $\Lambda$ at a specific order and $\Lambda$ at 4th order. The dots represent the average over a sample of 2000 tachyon-free models and the bars give the maximum deviation from this average.}}
\label{CoCConv}
\end{figure}

The practical way to perform this integral is as presented in \cite{D} using the expansion of the $\eta$ and $\theta$ functions in terms of the modular parameter, or more conveniently in terms of $q\equiv e^{2\pi i \tau}$ and $\bar{q}\equiv qe^{-2\pi i \bar{\tau}}$. This leads to a series expansion of the one-loop partition function which converges quickly as demonstrated in Figure \ref{CoCConv}.
The details and conventions for the q-expansions can be found in Appendix \ref{Appendix}.  All terms in the partition function sum (\ref{ZInt}) are modular functions of the variable $\tau$ and so we can rewrite the expression in terms of a q-expansion as
\begin{equation}
    Z = \sum_{n.m} a_{mn} \int_\mathcal{F} \frac{d^2\tau}{\tau_2^3} \, q^m \bar{q}^n = \sum_{n.m} a_{mn} \int_\mathcal{F} \frac{d^2\tau}{\tau_2^3} \,  e^{-2\pi\tau_2(m+n)} e^{2\pi i \tau_1(n-m)}.
    \label{QPF}
\end{equation}
It is important to note, that in the expression above, the $a_{mn}$ physically represent the difference between bosonic and fermionic degrees of freedom at each mass level, i.e. $a_{mn} = N_b - N_f$. Since the fundamental domain  $\mathcal{F}$ is symmetric with respect to $\tau_1$, only the even part of the $\tau_1$ exponential will contribute giving
\begin{equation} 
Z = \sum_{n.m} a_{mn} \int_\mathcal{F} \frac{d^2\tau}{\tau_2^3} \,  e^{-2\pi\tau_2(m+n)} \cos(2\pi\tau_1(m-n)) \eqqcolon \sum_{m,n} a_{mn} I_{mn}.
\label{QCoC}
\end{equation}
The integral over $\tau_1$ can be done analytically while the $\tau_2$ integral has to be done numerically. The analytic integral is calculated by splitting $\mathcal{F}$ into the two regions
\begin{align*}
\begin{cases}\mathcal{F}_1 = \{\tau\in\mathbb{C}\,|\,\tau_2\geq1 \;\land\; |\tau_1|<1/2\} \\
\mathcal{F}_2 = \{\tau\in\mathbb{C}\,|\,|\tau|^2>1 \;\land\;\tau_2<1\;\land\; |\tau_1|<1/2\},
\end{cases}
\end{align*} 
such that $\mathcal{F} =\mathcal{F}_1 \cup \mathcal{F}_2$. Performing the integration over $\tau_2$ in this way also gives insight into what terms can and cannot contribute to the partition function. The integral over $\mathcal{F}_2$ is always finite however, the integral over $\mathcal{F}_1$ diverges for specific values of $m,n$. We specifically find that the following cases arise:
\begin{align}
    I_{mn} = 
    \begin{cases}
    \infty \quad &\text{if} \quad m+n<0 \; \land \; m-n\notin\mathbb{Z}\backslash\{0\}\\
    \text{Finite} \quad &\text{Otherwise}.
    \end{cases}
    \label{I}
\end{align}
 The numerical values of the integrals $I_{mn}$ can be found in Table \ref{ITab} of Appendix \ref{Appendix}. We learn that as expected on-shell tachyonic states, i.e. states with $m=n<0$, have an infinite contribution. On the other hand, it is important to note that some off-shell tachyonic states may contribute a finite value to the partition function. The above result also shows that not only on-shell tachyonic states can cause a divergence, but some off-shell tachyonic states as well. These states, however, do not arise due to the modular invariance constraints on the coefficients $\CC{\alpha}{\beta}$, which only allows states with $m-n\in \mathbb{Z}$.

\begin{figure}[t]
\centering
\includegraphics[width=0.8\linewidth]{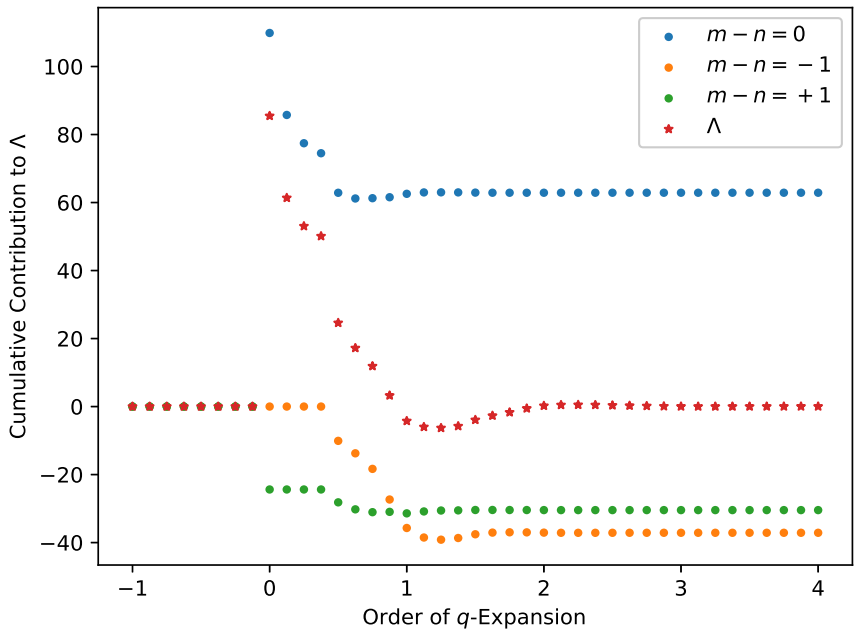}
\caption{\emph{A comparison of different contributions to $\Lambda$ for a model with $\Lambda = 0.03$ arranged as in (\ref{amn}). We see that the large positive contributions of the on-shell states are compensated by the negative contributions of the off-shell states.}}
\label{CoCContributions}
\end{figure}

The modular invariance constraint $m-n\in \mathbb{Z}$ means that the q-expansion of the partition function (\ref{QPF}) neatly arranges into the form
\begin{equation} 
a_{mn} =
\begin{pmatrix}
  0 & 0 & a_{-\frac{1}{2}-\frac{1}{2}} & 0 & 0 & 0 & a_{-\frac{1}{2}\frac{1}{2}} & 0 & 0 & 0  \\
  0 & 0 & 0 & a_{-\frac{1}{4}-\frac{1}{4}} & 0 & 0 & 0 & a_{-\frac{1}{4}\frac{3}{4}} & 0 & 0  \\
  a_{{0}{-1}} & 0 & 0 & 0 & a_{{0}{0}} & 0 & 0 & 0 & a_{{0}{1}} & 0  \\
  0 & a_{\frac{1}{4}-\frac{3}{4}} & 0 & 0 & 0 & a_{\frac{1}{4}\frac{1}{4}} & 0 & 0 & 0 & \ddots  \\
  0 & 0 & a_{\frac{1}{2}-\frac{1}{2}} & 0 & 0 & 0 & a_{\frac{1}{2}\frac{1}{2}} & 0 & 0 & 0  \\
  0 & 0 & 0 & a_{\frac{3}{4}-\frac{1}{4}} & 0 & 0 & 0 & a_{\frac{3}{4}\frac{3}{4}} & 0 & 0  \\
  a_{{1}{-1}} & 0 & 0 & 0 & a_{{1}{0}} & 0 & 0 & 0 & a_{{1}{1}} & 0  \\
  0 & \ddots & 0 & 0 & 0 & \ddots & 0 & 0 & 0 & \ddots  \\
\end{pmatrix} 
\label{amn}
\end{equation}
i.e. into series of states with $n-m = p\in \mathbb{Z}$. This gives a convenient way to examine the different contributions to the cosmological constant (\ref{QCoC}) and compare the effect of on and off-shell states. As an example we consider a model with a small value for the cosmological constant as shown Figure \ref{CoCContributions}.
We see that the suppressed value of $\Lambda$ is due to the cancellation between the large positive contributions from the on-shell states and the negative contributions from the off-shell states. Indeed, in general we find that for our set of models, the only positive contributions to $\Lambda$ come from on-shell states and so these states can give us a handle on the expected value of the cosmological constant.

As we have seen in Figure \ref{CoCConv}, for our tachyon-free models, $\Lambda$ always converges and does so rapidly starting from 2nd order in q. It is often stated that the finiteness of string theory is due to supersymmetry, which in our case is not present, thus one may wonder how the partition function of non-supersymmetric theories manages to remain finite. For supersymmetric free fermionic theories the usual $S$-vector (\ref{tendsvector}) ensures  that the bosonic and fermionic degrees of freedom are exactly matched at each mass level. That is, for a supersymmetric theory we necessarily have that $a_{mn} = 0$ for all $m$ and $n$, which in turn causes the vanishing of the cosmological constant as one expects. For our non-supersymmetric models, the lack of an $S$-vector means that such cancellations are not ensured and so such theories in general produce a non-zero value for $\Lambda$. It is, however, not obviously clear that they should produce finite partition functions. Such finiteness is achieved through a mechanism called misaligned supersymmetry as presented in \cite{MSUSY2,MSUSY,carlo}. 

\begin{figure}[t]
\centering
\includegraphics[width=0.8\linewidth]{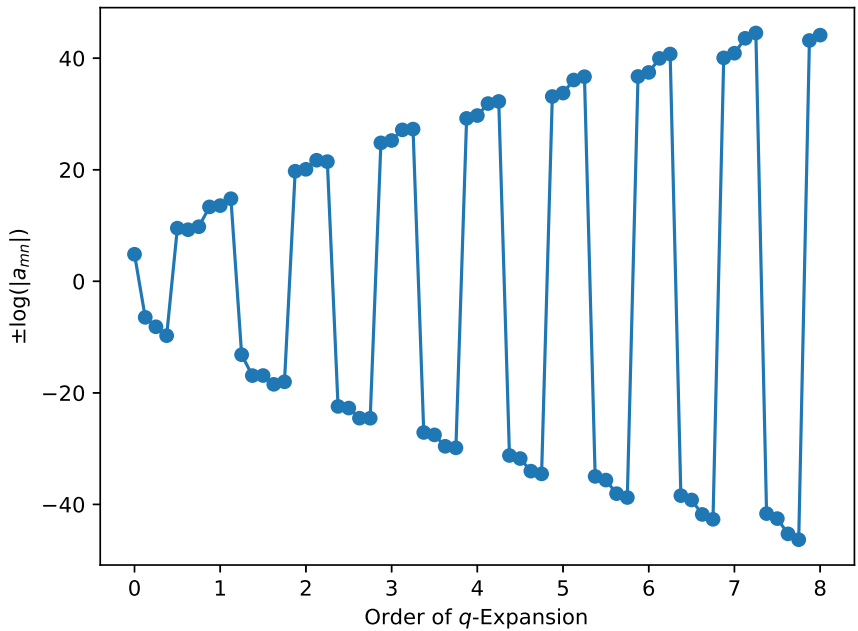}
\caption{\emph{The boson-fermion oscillation of misaligned supersymmetry for the on-shell states of one of our models to 8$^{th}$ order in the q-expansion. The overall sign of $\pm\log(\mid a_{mn}\mid)$ is chosen according to the sign of $a_{mn}$.}}
\label{BFOsc}
\end{figure}

As one expects, the degeneracy of states grows rapidly going up the infinite tower of massive states. This growth, in theory, could counteract the suppression received from the decreasing contributions from the integrals in Table \ref{ITab} and cause divergences. The mechanism of misaligned supersymmetry, however, causes the states in the massive tower to oscillate between an excess of bosons and an excess of fermions. This behaviour is referred to as boson-fermion oscillation. Our models indeed present this behaviour as shown in Figure \ref{BFOsc}.
Instead of cancelling level-by-level as in the supersymmetric case, the cancellation is misaligned causing the oscillation meaning a large positive contribution is followed by an even larger negative contribution and so on. This mechanism ensures that the partition function of our non-supersymmetric models remains finite.

\subsection{$N_b = N_f$ at the Massless Level} \label{NbNf}
The discussion above shows that while for non-supersymmetric theories there is no mechanism which ensures the vanishing of $a_{mn}$ at any allowed level, there is, however, nothing preventing it from happening. It is indeed possible to find models within our classification set-up detailed in Section \ref{construction} which have $a_{00}=0$, \textit{i.e. } $N_b^0 = N_f^0$.

In the analysis of the one-loop potential in \cite{FR}, no models are found which exhibit $N_b^0 = N_f^0$ at the free fermionic point in the sample explored. Instead they use techniques developed in \cite{F} to move away from the free fermionic point using an analogous orbifold rewriting of the partition function and find models with $N_b^0 = N_f^0$ at a generic point in the moduli space. In our analysis we stay at the free fermionic point and it turns out that we do find models with $N_b^0 = N_f^0$ and an example model is presented in Section \ref{exampleModel}. 

It is convenient to summarise the various contributions to $a_{00}$ in the form of Table \ref{a00Contribs}. We use the notation for sectors laid out in Section \ref{sectors}. For simplicity, and since we restrict our classification to models with no enhancements, the contributions of vector bosons from sectors $z_1,z_2,z_1+z_2$ are ignored.

\begin{center}
\small
\setlength{\tabcolsep}{2pt}
\begin{table}[!htb]
\centering
\begin{tabular}{|c|r||c|r|}
\hline
Sector&	$N_b-N_f$ & Sector&	$N_b-N_f$ \\
\hline
$\ket{NS}$&  $304$ & $\{ \bar{y}_{NS}^i/\bar{w}_{NS}^i\}
  \ket{\tilde{V}}$ & -8 \\ 
\hline
$\ket{B^{1,2,3}}$ & $-32$&$\delta^{1,...,30}$ & 16\\
\hline
$\ket{\tilde{x}}$ & -256 &$\{\bar{\psi}^{a(*)}/\bar{\eta}^{b(*)}\}\ket{\gamma^{1,...,15}}$ & 64 \\
\hline
$\{\bar{\psi}^{a(*)}\}\ket{V^{1,2,3}}$ & 32 &$\{\bar{y}_{NS}^i/\bar{w}_{NS}^i\}\ket{\gamma^{1,...,15}}$ & 4 \\
\hline
$\{\bar{\phi}^{\{1,2\}/\{3,4\}/\{5,6\}/\{7,8\}(*)}\}\ket{V^{1,2,3}}$ & $8$&$\{\bar{\phi}^{\{1,2\}/\{3,4\}/\{5,6\}/\{7,8\}(*)}\}\ket{\gamma^{1,...,15}}$ & 8\\ 
\hline
$\{\bar{y}^i/\bar{w}^i\}\ket{V^{1,2,3}}$ & 4 &$\{y_{NS}^i/w_{NS}^i\}\{\bar{y}_{NS}^j/\bar{w}_{NS}^j\}\ket{z_{1/2}}$ & 8\\ 
\hline
$\ket{H^{1,...,6}}$ & 16 & $\{y_{NS}^i/w_{NS}^i\}\{\bar{\eta}^{b(*)}\}\ket{z_{1/2}}$ & 32\\ 
\hline
$\ket{H^{7,...,18}}$ & -8 & $\{y_{NS}^i/w_{NS}^i\}\{\bar{\phi}^{\{5,6,7,8\}/ \{1,2,3,4\}(*)}\}\ket{z_{1/2}}$ & 16\\
\hline
$\{ \bar{\psi}^{1,...,5(*)},\bar{\eta}^{1,2,3(*)},\bar{\phi}_{NS}^{(*)}\}
  \ket{\tilde{V}}$ & -192 & $\{y_{NS}^i/w_{NS}^i\}\ket{z_{1}+z_2}$ & 8\\ 
\hline  

\end{tabular}
\caption{\label{a00Contribs} \emph{Contributions of massless sectors to $a_{00}$ when present in Hilbert space of a model. As noted $a_{00}=N_b^0-N_f^0$, so bosonic contributions are positive and fermionic are negative. The superscripts used here are $i\neq j =1,...,6$, $a=1,...,5$ and $b=1,2,3$. The NS subscript means that the oscillator has Neveu-Schwarz boundary conditions in the sector.}}
\end{table}
\end{center}

\clearpage

\section{Results of Classification} \label{results}
Having discussed how to determine key features of the massless spectrum and
how to calculate the partition function and cosmological constant for
our \ds -models we can now present some statistics derived from a sample in the space of models. As mentioned in Section \ref{construction},
the space of all models is $2^{66}\sim 10^{19.9}$ and so a complete
classification is far beyond the computing power at our disposal.
Instead, we explore a sample of $2\times 10^9$ models of which only
around 1 in 185 are tachyon-free that we take forward for further analysis. We will start with some results of key aspects of the massless spectrum. 
\subsection{Results from Massless Spectrum}
\begin{figure}[!ht]
\centering
\includegraphics[width=0.8\linewidth]{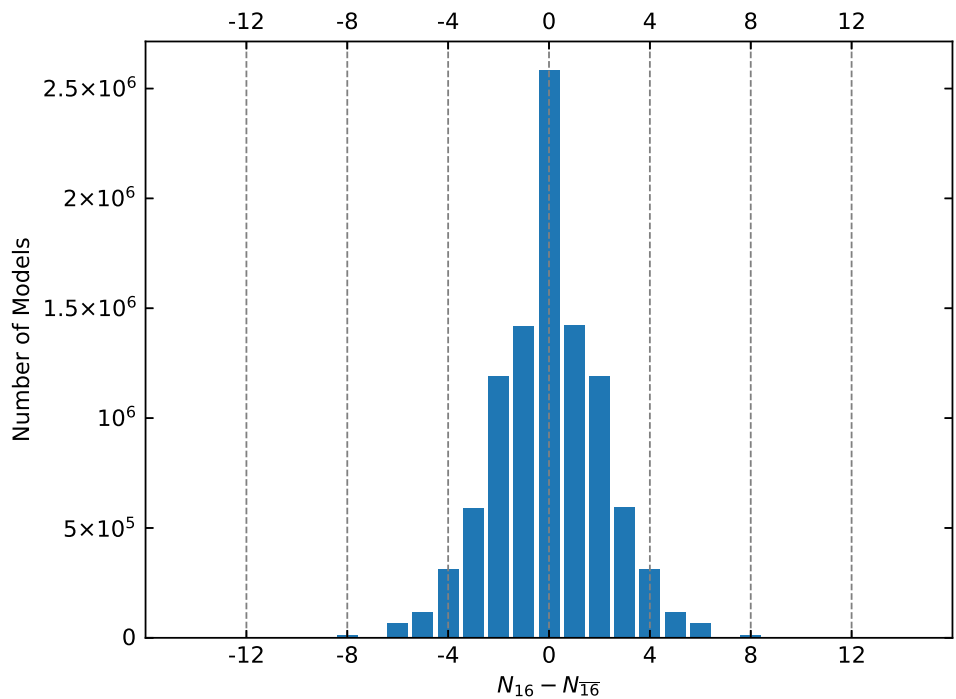}
\caption{\label{NetChiralGraph}\emph{Number of models versus net chiral generations from a random sample of $10^7$ tachyon free $SO(10)$ models.}}
\end{figure}

\begin{figure}[!hbt]
\centering
\includegraphics[width=0.8\linewidth]{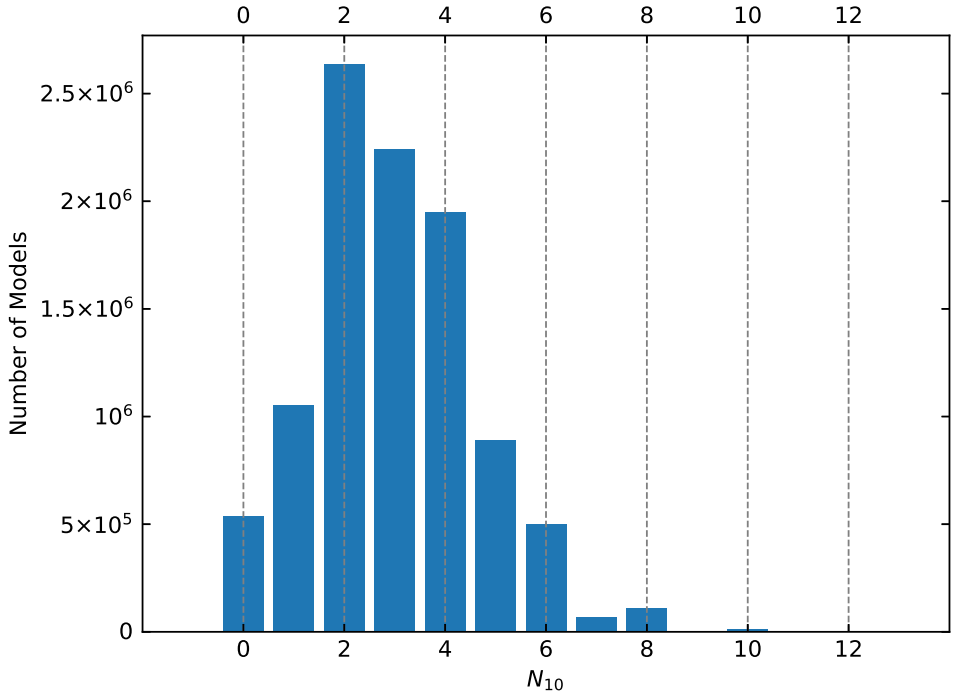}
\caption{\label{Num10sGraph}\emph{Number of models versus number of vectorial
    $\mathbf{10}$ sectors from a random sample of $10^7$ tachyon free $SO(10)$
    models.}}
\end{figure}

From our sample of $2\times 10^9$ models we choose $10^7$ tachyon-free
models and display the results for their $SO(10)$ observable representations.
In Figure \ref{NetChiralGraph} the net chirality, $N_{16}-N_{\overline{16}}$,
distribution is displayed and in Figure \ref{Num10sGraph} the distribution
of their number of vectorial $\mathbf{10}$ representations is displayed.  
The familiar normal distribution also found in all other classifications for the supersymmetric cases is uncovered. This is hardly surprisingly since the structure of the fermionic $\mathbf{16}/\overline{\mathbf{16}}$ is unchanged for our models. 
From Figure \ref{Num10sGraph} we see that the large majority of models
contain at least 1 vectorial $\mathbf{10}$ which may be used to generate a
bidoublet Higgs representation when the $SO(10)$ is broken. 

In order to see more clearly the statistics from our $2\times 10^9$
sample we display the frequency of $SO(10)$ models as several
phenomenological constraints are considered in Table \ref{Statstable}.
\begin{table}[!htb]
\centering
\begin{tabular}{|c|l|r|c|r|}
\hline
&Constraints & \parbox[c]{2.5cm}{Total models in sample}& Probability \\
\hline
 & No Constraints & $2\times 10^9$ & $1$  \\ \hline
(1)&{+ Tachyon-Free} & 10741667 & $5.37\times 10^{-3}$  \\  \hline
(2)& {+ No Observable Enhancements} & 10741667 & $5.37\times 10^{-3}$  \\ \hline
(3)&{+ No Hidden Enhancements} & 9921843 & $4.96\times 10^{-3}$  \\  \hline
(4)&{+ $N_{16}-N_{\overline{16}}\geq 6$} & 69209 & $3.46\times 10^{-5}$  \\  \hline
(5)&{+ $N_{10}\geq 1$}& 69013 & $3.45\times 10^{-5}$ \\ \hline
(6)&{+ $a_{00}=N_b^0-N_f^0=0$}& 3304&$1.65\times 10^{-6}$ \\ \hline
\end{tabular}
\caption{\label{Statstable} \emph{Phenomenological statistics from sample of $2\times 10^9$ $SO(10)$ \ds -models.}}
\end{table}

These results confirm the observation made in previous sections that
there are no tachyon-free models in our construction which have observable
enhancements. In phenomenological terms, we do not need to worry about
enhancements of the hidden sector gauge group, but they are included in the table for
completeness. The next constraints we add are much like the
so-called `fertility constraints' implemented in \cite{frs,lrsfertile}. The constraint
on the net chirality $N_{16}-N_{\overline{16}} \geq 6$ is a necessary,
but not sufficient, condition for the existence of 3 or more chiral
generations at the level of the standard model.
The condition $N_{10}\geq 1$ ensures at least one state exists that
can produce a Standard Model Higgs doublet and can be used to break the
electroweak symmetry. 
Finally, we implement a
condition on the q-expansion coefficient $a_{00}=0$ which corresponds to
finding models with $N_b=N_f$ at the massless level as discussed in
Section \ref{PF}. 

The $3304$ 
models satisfying all these constraints are notable, particularly in
regard to this final condition of $N_b^0=N_f^0$. 
Inspecting the patterns
in the spectra of these $3304$ models revealed that
$\sim 58\%$ 
contain the vector $\tilde{x}$ in their spectrum. In these cases the
large negative contribution of $-256$ that $\tilde{x}$ contributes to
$a_{00}$ is helpful in ensuring $N_b^0=N^f_0$. Of those models not containing $\tilde{x}$ $\sim 70\%$ obtained the  
large negative contribution of $-192$ from one of the additional
vectorials $\tilde{V}=\tilde{S},\tilde{S}+z_1,\tilde{S}+z_2,\tilde{S}+z_1+z_2$
with mixed charges under the observable and hidden groups, i.e. the sectors
$\{ \bar{\psi}^{1,...,5},\bar{\eta}^{1,2,3},\bar{\phi}_{NS}\}\ket{\tilde{V}}$. Again this large negative contribution helps in matching the number of massless fermions to massless bosons. 

\subsection{Results for Cosmological Constant and $N^0_b-N^0_f$}
As the value of the constant term $a_{00} = N_b^0-N_f^0$ and the cosmological constant $\Lambda$ vary from model-to-model, it is interesting to see what range of values these non-supersymmetric models can produce. 

\begin{figure}[t]
\centering
\includegraphics[width=0.85\linewidth]{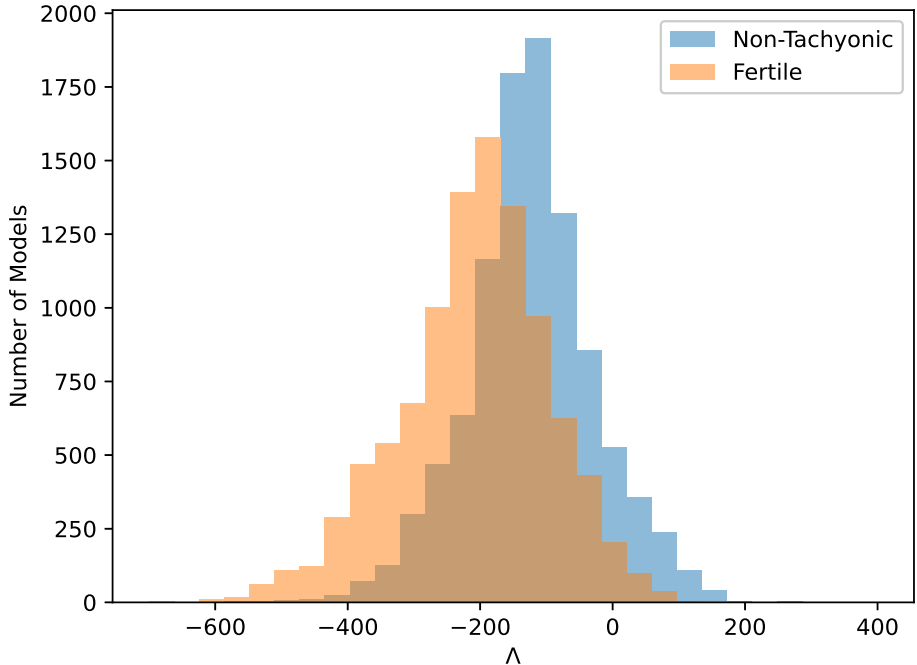}
\caption{\emph{The distribution of the cosmological constant for a sample of $10^4$ non-tachyonic and $10^4$ fertile models models.}}
\label{LDist}
\end{figure}

The distribution of the cosmological constant $\Lambda$ is shown in in Figure \ref{LDist}, for a sample of $10^4$ non-tachyonic and $10^4$ fertile models. By non-tachyonic we mean that only condition (1) of Table \ref{Statstable} is satisfied, while fertile models satisfy all conditions (1)-(5). It is important to note that values presented in Figure \ref{LDist} are at the special free fermionic point in moduli space. This means that moving away from this point will change these values and if there are unfixed moduli, there is nothing preventing this from happening. This is indeed the case for our class of models. 

\begin{figure}[t]
\centering
\includegraphics[width=0.85\linewidth]{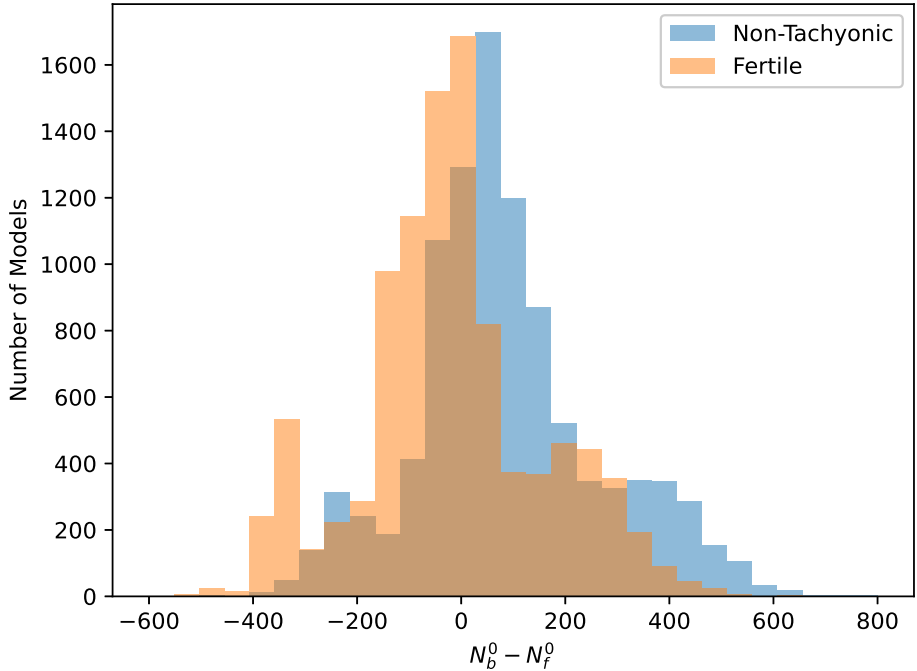}
\caption{\emph{The distribution of the constant term $a_{00} = N_b^0-N_f^0$ for a sample of $10^4$ non-tachyonic and $10^4$ fertile models.}}
\label{CDist}
\end{figure}

Another interesting quantity in the partition functions is boson-fermion degeneracy at the massless level. As discussed in Section \ref{PF}, the on-shell states provide the majority of positive contributions to the partition function, the largest of which is the massless term. Thus the value of $a_{00}$ gives a good handle on the value of the cosmological constant. 
It is also, of course, interesting for the discussion of phenomenological features and stability as explained in Section \ref{NbNf}. The distribution of values of $a_{00} = N_b^0-N_f^0$ for a sample of $10^4$ non-tachyonic and $10^4$ fertile models is shown in Figure \ref{CDist}.

From Figures \ref{LDist} and \ref{CDist} we see that the fertility conditions have a measurable effect on the distribution of $\Lambda$ and $a_{00}$, that is, they slightly shift the values of both to the negative. This is an interesting effect and is likely due to condition (4) in Table \ref{Statstable}. Even though the fertility condition (4) is directed at ensuring the difference $N_{16}-N_{\overline{16}}$ is greater than 6, in doing this it also results in fertile models having a larger average total $N_{16}+N_{\overline{16}}$ compared to non-fertile models. As specified in Table \ref{a00Contribs}, these sectors contribute a value of $-32$ to $a_{00}$ and thus appear to cause the shift toward smaller values for $a_{00}$ and as a consequence also for $\Lambda$.

\subsection{A Model with $N_b^0=N_f^0$} \label{exampleModel}

From the $3304$ fertile models with $N_b^0=N_f^0$ we present an analysis of
the key features of the massless spectrum for one example model, as well as
presenting its partition function and cosmological constant. The model we
choose has the GGSO matrix
{\begin{equation}
\small
\CC{v_i}{v_j}= 
\begin{blockarray}{ccccccccccccc}
&\mathbf{1}& \tilde{S} & e_1 & e_2 & e_3 & e_4 & e_5 & e_6 & b_1 & b_2&b_3&z_1 \\
\begin{block}{c(rrrrrrrrrrrr)}
\mathbf{1}&-1& -1&  1& -1& -1& -1&  1&  1&  1&  1& -1& -1\ \\
\tilde{S}& -1&  1& -1& -1& -1&  1&  1&  1&  1& -1& -1&  1\ \\
e_1&  1& -1& -1& -1& -1& -1& -1& -1& -1& -1& -1&  1\ \\
e_2& -1& -1& -1&  1& -1&  1&  1& -1&  1& -1&  1& -1\ \\
e_3& -1& -1& -1& -1&  1& -1& -1&  1&  1&  1& -1& -1\ \\
e_4& -1&  1& -1&  1& -1&  1&  1&  1&  1& -1&  1&  1\ \\
e_5&  1&  1& -1&  1& -1&  1& -1&  1& -1&  1&  1& -1\ \\
e_6&  1&  1& -1& -1&  1&  1&  1& -1& -1&  1&  1&  1\ \\
b_1&   1& -1& -1&  1&  1&  1& -1& -1&  1& -1&  1&  1\ \\
b_2 & 1&  1& -1& -1&  1& -1&  1&  1& -1&  1& -1&  1\ \\
b_3& -1&  1& -1&  1& -1&  1&  1&  1&  1& -1& -1& -1\ \\ 
z_1&-1& -1&  1& -1& -1&  1& -1&  1&  1&  1& -1& -1\ \\ 
\end{block}
\end{blockarray}
\end{equation}}
This model has $N_{16}=7$, $N_{\overline{16}}=1$ and $N_{10}=8$ and thus
satisfies the constraints imposed in Table \ref{Statstable}.
Furthermore, in this model the  $\tilde{x}$--sector produces massless states,
which in the supersymmetric models would correspond to the
presence of the $S+x$ sector when $x$ enhances the $SO(10)$ symmetry
to $E_6$.
In that case, the $S+x$ include the superpartners of 
the gauge vector bosons of the sector $x$,
i.e. the gauginos.
So in this case,
we have the gauginos but not the vector bosons.


Our model also contains 6 bosonic hidden states from the sectors $H^{1,...,6}$ and 48 fermionic hidden states from the $H^{7,...,18}$. There are additional vectorials from the sectors $e_3+e_4+e_5+e_6, \ e_1+e_2+e_3+e_6$ and $e_1+e_2+e_3+e_4$ with observable oscillators $\{\bar{\psi}^a,\bar{\eta}^{b}\}$, $a=1,...,5, \ b=1,2,3$ which cannot couple with observable states from $B^{1,2,3}$ since it cannot conserve the charges of the $U(1)_{1,2,3}$ in particular. However, these three sectors may provide couplings at higher order. 

The partition function is calculated in terms of its $q$-expansion and so it can be specified by a matrix of coefficients $a_{mn}$ as in (\ref{amn}). For our example model these values are presented in Table \ref{QEx}.
\begin{table}[!htb]
\centering
\setlength{\tabcolsep}{2pt}
\footnotesize
\begin{tabular}{|r||r|r|r|r|r|r|r|r|r|r|r|r|r|r|r|r|}
\hline
     & -1 &-7/8&-3/4&-5/8&-1/2 & -3/8& -1/4& -1/8 & 0     & 1/8  & 1/4   & 3/8   & 1/2  & 5/8   & 3/4   & 7/8    \\\hhline{|=|=|=|=|=|=|=|=|=|=|=|=|=|=|=|=|=|}
-1/2 & 0  & 0  & 0  & 0  & 0   & 0   & 0   & 0    & 0     & 0    & 0     & 0     & 320  & 0     & 0     & 0      \\\hline
-3/8 & 0  & 0  & 0  & 0  & 0   & 0   & 0   & 0    & 0     & 0    & 0     & 0     & 0    & 896   & 0     & 0      \\\hline
-1/4 & 0  & 0  & 0  & 0  & 0   & 0   & 0   & 0    & 0     & 0    & 0     & 0     & 0    & 0     & 5696  & 0      \\\hline
-1/8 & 0  & 0  & 0  & 0  & 0   & 0   & 0   & 0    & 0     & 0    & 0     & 0     & 0    & 0     & 0     & 29312  \\\hline
0    & 2  & 0  & 0  & 0  & 0   & 0   & 0   & 0    & 0     & 0    & 0     & 0     & 0    & 0     & 0     & 0      \\\hline
1/8  & 0  & 0  & 0  & 0  & 0   & 0   & 0   & 0    & 0     & -288 & 0     & 0     & 0    & 0     & 0     & 0      \\\hline
1/4  & 0  & 0  & 0  & 0  & 0   & 0   & 0   & 0    & 0     & 0    & -4512 & 0     & 0    & 0     & 0     & 0      \\\hline
3/8  & 0  & 0  & 0  & 16 & 0   & 0   & 0   & 0    & 0     & 0    & 0     & -9808 & 0    & 0     & 0     & 0      \\\hline
1/2  & 0  & 0  & 0  & 0  & 224 & 0   & 0   & 0    & 0     & 0    & 0     & 0     & 1344 & 0     & 0     & 0      \\\hline
5/8  & 0  & 0  & 0  & 0  & 0   & 416 & 0   & 0    & 0     & 0    & 0     & 0     & 0    & 36640 & 0     & 0      \\\hline
3/4  & 0  & 0  & 0  & 0  & 0   & 0   & 576 & 0    & 0     & 0    & 0     & 0     & 0    & 0     & 78080 & 0      \\\hline
7/8  & 0  & 0  & 0  & 0  & 0   & 0   & 0   & -320 & 0     & 0    & 0     & 0     & 0    & 0     & 0     & 212928 \\\hline
1    & 32 & 0  & 0  & 0  & 0   & 0   & 0   & 0    & -1440 & 0    & 0     & 0     & 0    & 0     & 0     & 0      \\\hline
\end{tabular}
\caption{\emph{The $q$-expansion of the partition function for our example model. Each entry in the table represents the coefficient $a_{mn}$ in the partition function sum (\ref{QPF}), with the first column and row being the mass levels for the left and right moving sectors respectively.}}
\label{QEx}
\end{table}
We see that indeed this model has $a_{00} = N_b^0-N_f^0 = 0$ as advertised and the series of states arrange according to (\ref{amn}). The absence of on-shell tachyons is explicit and the contribution from off-shell tachyonic states is non-zero as expected. We also find that the consistency condition $a_{0-1}=2$ for the proto-graviton (\ref{protograviton}) as described in \cite{ADM,D} is also satisfied. 

The cosmological constant can also be calculated according to (\ref{QCoC}) with the modular integral quickly converging after 2$^{nd}$ order in $q$. In this case it takes the value 
\begin{equation}
    \Lambda = \sum_{m,n} a_{mn} I_{mn} = -149.77
\end{equation}
at the free fermionic point. As we see it is negative which is the case for most models with $N_b^0 = N_f^0$. This is due to the fact that the largest positive contributions to the partition function come from the light on-shell states and in particular from the massless states. If $N_b^0 - N_f^0 =0$, this is zero and the negative contributions from the light off-shell tachyons produce a negative value for $\Lambda$. This is indeed the case for all $3304$ such models in our scan.

\section{Discussion and Conclusion}\label{conclusion}

In this paper we developed systematic computerised tools to classify large
spaces of free fermion heterotic string vacua that correspond to
compactifications of ten dimensional tachyonic vacua. From the
point of view of the four dimensional constructions this is
achieved by the general \ds--map. Our previous
\unahe--based model \cite{stable} was similarly constructed
from the model published in \cite{cfmt}, which raises the question
what are the consequences of applying the map to generic models, {\it i.e.}
what are the relations between the spectra of the two mapped models, and
what are the general patterns. This relation is similar to the
general relation exhibited by the spinor--vector
duality map, and the two may in fact be manifestation of a much larger
symmetry structure \cite{panos}.

Adopting the classification methodology developed for supersymmetric
free fermionic models entails the proliferation of tachyon producing
sectors in the \ds--mapped models. The systematic classification therefore
requires detailed analysis of these sectors that was discussed in Section
\ref{tsa}. In the analysis of the massless sectors separate attention
to bosonic and fermionic sectors is required and was discussed in Section
\ref{sectors}. In Section \ref{PF} we discussed the general analysis of the
partition function and its $q$--expansion in left and right moving
energy modes. The analysis of the partition function is
particularly instrumental in the case of non--supersymmetric
string vacua as it gives a direct handle on the physical states at different
mass levels. Of particular interest in the $q$--expansion is
the $a_{00}=N_b^0-N_f^0$ term, which counts the difference between
massless bosons and fermions in the spectrum of the string vacuum.
In supersymmetric models the number of fermionic and bosonic
degrees of freedom are matched at all mass levels, and hence the
partition function and the vacuum energy are identically zero.
In non--supersymmetric models there is a generic mismatch at different
mass levels, which is partially compensated by the so--called misaligned
supersymmetry \cite{MSUSY}. It has been argued that in tachyon--free
non--supersymmetric models with $a_{00}=0$ the vacuum energy may
be suppressed by the volume of the compactified dimensions \cite{kp}.

In Section \ref{results} we presented the results of the classification
of the order of $2\times 10^9$ random GGSO phases that generate the space
of vacua spanned by the basis vectors in eq. (\ref{basis})
and the 66 independent one--loop GGSO phases. The analysis reveals
that tachyon--free models occur with $\sim 5\times 10^{-3}$ probability. Furthermore, we analysed this data by further imposing some fertility conditions $N_{16}-N_{\overline{16}}\geq 6$ and $N_{10}\geq 1$ and found fertile models with $a_{00}=0$ with frequency $\sim 2\times 10^{-6}$ in our sample. In Figures \ref{CDist} and \ref{LDist} a notable shift in values of the cosmological constant and the $a_{00}$ term were detected for fertile models compared with a random sample of non-tachyonic vacua. 

These results reveal that extracting interesting
phenomenological models necessitates the development of more sophisticated
computerised methods than the random generation method. This is particularly
true in light of the fact that generating viable symmetry breaking pattern
may necessitate breaking the $SO(10)$ symmetry to the Standard Model subgroup.
The \ds--map entails that scalar degrees of freedom in the spinorial
sixteen representation of $SO(10)$ are shifted to the massive spectrum.
The consequence is that the spectrum does not contain the neutral component
in the $16$ of $SO(10)$ required to break the remnant unbroken gauge symmetry
down to the Standard Model gauge group. The only available states are
exotic states that carry fractional $U(1)_{Z^\prime}$ charge and appear
in the heterotic string Standard--like Models \cite{exotics}. 
This assertion requires of course further investigation that will be
scrutinised in future work. The lesson may be that quasi--realistic models
in this class may only be possible for a very restricted and narrow
set of models, rather than the more generic set, which is the prevalent
experience with supersymmetric constructions.
In forthcoming work these questions are investigated in tachyon--free
Pati--Salam models, including the inclusion of fertility conditions.
The increased space of vacua, in particular in the case of Standard--like
models, requires adaptation of novel computational techniques \cite{fhpr}.

Following from or previous paper \cite{stable}
the analysis and results presented in this work open up new vistas
in string phenomenology. It reveals the potential relevance of string
vacua that have been previously considered to be irrelevant. The number
of questions to explore is large and may potentially
provide insight into some of the prevailing problems in string phenomenology.
Interpolation between the supersymmetric vacua and our tachyon--free constructions, as well as with the two dimensional MSDS constructions
\cite{msds}, 
may shed some light on the problem of supersymmetry breaking
and vacuum energy in string
theory. This can be carried out in a subset of the basis
vectors {\it e.g.} $\{1,{\tilde S}, b_1\}$ or $\{1,{\tilde S}, {\tilde x}\}$.
Another question of interest is the question of stability of the
tachyon--free models. This question is necessarily tied with the
non--vanishing one--loop vacuum energy in these models. In this
respect it will be interesting to analyse the one--loop diagram
that arises in these models due to the existence of an anomalous
$U(1)$ symmetry \cite{ads}
and to examine whether two diagrams can be
cancelled against each other. Finally, further understanding
of the symmetries that underlie the partition function at all mass levels,
as exhibited at the massless level by the \ds\,  and $\tilde{x}$ maps,
are important to extract. 

\section*{Acknowledgments}

The work of VGM is supported in part by EPSRC grant EP/R513271/1. The work of BP is supported in part by STFC grant ST/N504130/1.

\newpage

\appendix

\section{Theta Functions and Integrals} \label{Appendix}

As we have seen, the partition functions are given in terms of Jacobi theta and Dedekind eta functions. These are the functions that we have to expand in terms of the parameters $q\equiv e^{2\pi i \tau}$ and $\bar{q}\equiv qe^{-2\pi i \bar{\tau}}$ in order to perform the integrals over the modular domain. Further details can be found in \cite{theta}, and for completeness we  list below the q-expansions used to calculate the partition function.

The q-expansions of the $\theta$ and $\eta$ functions are easily derived form their definitions   
\begin{align}
\vartheta_2 =& \sum_{n \in \mathbb{Z}} q^{(n+1 / 2)^{2} / 2}  = 2q^{1/8} + 2q^{9/8} + 2q^{25/8} + \cdots \label{t2}\\
\vartheta_3 =& \sum_{n \in \mathbb{Z}} q^{n^{2} / 2}  = q^0 + 2q^{1/2} + 2q^2 + \cdots \label{t3}\\
\vartheta_4 =& \sum_{n \in \mathbb{Z}}(-1)^{n} q^{n^2 / 2}  = q^0 - 2q^{1/2} + 2q^2 + \cdots\label{t4}\\
\eta =& \, q^{1 / 24} \prod_{n=1}^{\infty}\left(1-q^{n}\right) = q^{1/24} - q^{25/24} - q^{49/24} + \cdots .
\end{align}
Since the partition function only involves negative powers of $\eta$, the above expression is not useful in practical terms. One can, however, find a general expansion for $\eta^{-1}$ using the multinomial theorem which gives
\begin{equation}
\eta^{-1} = q^{-1 / 24} \prod_{n=1}^{\infty} \sum_{k=0}^{\infty}\binom{-1}{k}(-1)^{k} q^{nk}=q^{-1 / 24}+q^{23 / 24}+ 2 q^{47/24} + \cdots.
\end{equation}

Substituting the above q-expansions into the partition function we arrive at the form stated in (\ref{QPF}) and thus all that remains is to calculate the integrals of the form
\begin{equation}
    I_{mn} = \int_\mathcal{F} \frac{d^2\tau}{\tau_2^3} \, q^m \bar{q}^n = \int_\mathcal{F} \frac{d^2\tau}{\tau_2^3} \,  e^{-2\pi\tau_2(m+n)} e^{2\pi i \tau_1(n-m)}.
\end{equation}
As described above in Section \ref{PF}, this is done using both analytic an numerical techniques. The values of these integrals can be found in Table \ref{ITab}, where they are listed for the range $m,n\leq 1$.

\begin{sidewaystable}
    \setlength{\tabcolsep}{3pt}
    \centering
    \begin{tabular}{|r|l|l|l|l|l|l|l|l|l|}
    \hline
        $q\backslash \bar{q}$ & -1     & - 3/4 & - 1/2 & - 1/4 & 0     &  1/4 &  1/2 &  3/4 & 1     \\ \hline
        -1     & $\infty$ & $\infty$ & $\infty$ & $\infty$ & -1.22$\times 10^1$ & $\infty$ & $\infty$ & $\infty$ & 9.90$\times 10^{-3}$ \\ \hline
        - 3/4 & $\infty$ & $\infty$ & $\infty$ & $\infty$ & $\infty$ & -6.17$\times 10^{-1}$ & $\infty$ & -1.34$\times 10^{-1}$ & -8.55$\times 10^{-3}$ \\ \hline
        - 1/2 & $\infty$ & $\infty$ & $\infty$ & $\infty$ & $\infty$ & $\infty$ & -3.15$\times 10^{-2}$ & -1.78$\times 10^{-2}$ & -2.98$\times 10^{-3}$ \\ \hline
        - 1/4 & $\infty$ & $\infty$ & $\infty$ & $\infty$ & $\infty$ & 3.35$\times 10^{-1}$ & 1.30$\times 10^{-2}$ & -1.63$\times 10^{-3}$ & -6.47$\times 10^{-4}$ \\ \hline
        0     & -1.22$\times 10^1$ & $\infty$ & $\infty$ & $\infty$ & 5.49$\times 10^{-1}$ & 5.56$\times 10^{-2}$ & 5.61$\times 10^{-3}$ & 2.25$\times 10^{-4}$ & -8.46$\times 10^{-5}$ \\ \hline
         1/4 & $\infty$ & -6.17$\times 10^{-1}$ & $\infty$ & 3.35$\times 10^{-1}$ & 5.56$\times 10^{-2}$ & 1.00$\times 10^{-2}$ & 1.54$\times 10^{-3}$ & 1.70$\times 10^{-4}$ & 3.21$\times 10^{-6}$ \\ \hline
         1/2 & $\infty$ & $\infty$ & -3.15$\times 10^{-2}$ & 1.30$\times 10^{-2}$ & 5.61$\times 10^{-3}$ & 1.54$\times 10^{-3}$ & 3.30$\times 10^{-4}$ & 5.52$\times 10^{-5}$ & 6.05$\times 10^{-6}$ \\ \hline
         3/4 & $\infty$ & -1.34$\times 10^{-1}$ & -1.78$\times 10^{-2}$ & -1.63$\times 10^{-3}$ & 2.25$\times 10^{-4}$ & 1.70$\times 10^{-4}$ & 5.52$\times 10^{-5}$ & 1.29$\times 10^{-5}$ & 2.22$\times 10^{-6}$ \\ \hline
        1     & 9.90$\times 10^{-3}$ & -8.55$\times 10^{-3}$ & -2.98$\times 10^{-3}$ & -6.47$\times 10^{-4}$ & -8.46$\times 10^{-5}$ & 3.21$\times 10^{-6}$ & 6.05$\times 10^{-6}$ & 2.22$\times 10^{-6}$ & 5.47$\times 10^{-7}$ \\ \hline
    \end{tabular}
    \caption{\emph{The values of the integral $I_{mn}$ for $m,n\leq 1$. The first column and row denotes the value of $m$ and $n$ respectively. }} 
    \label{ITab}
\end{sidewaystable}

\clearpage

\bibliographystyle{unsrt}

\end{document}